\documentclass[article,longauth]{aa}
\usepackage{natbib}
\usepackage{longtable,lscape}
\usepackage{amsmath}
\usepackage{natbib}
\bibpunct{(}{)}{;}{a}{}{,}
\usepackage{graphicx}
\usepackage{multirow}

\usepackage{graphicx}
\usepackage{natbib}
\usepackage[colorlinks=true,citecolor=blue]{hyperref}
\usepackage{color}
\usepackage{xspace}
\usepackage{dcolumn}
\usepackage{longtable}
\usepackage{lscape}
\usepackage{soul}
\usepackage{wasysym}
\usepackage{rotating}
\usepackage{afterpage}
\usepackage{mathtools}
\usepackage{blindtext}

\newcommand{\MJup}{M$_{\mathrm{Jup}}$\xspace}
\newcommand{\RJup}{R$_{\mathrm{Jup}}$\xspace}

\newcommand{\MSun}{M$_{\odot}$\xspace}

\newcommand{\rcra}{R\,CrA}

\newcommand{\mic}{$\mu$m\xspace}
\newcommand{\as}{\hbox{$^{\prime\prime}$}\xspace}

\usepackage[varg]{txfonts}
\begin{document}
\title{Exploring the \rcra\ environment with SPHERE:} 
\subtitle{Discovery of a new stellar companion\thanks{Based on observations made
    with European Southern Observatory (ESO) telescopes at Paranal Observatory
    in Chile, under programs ID 095.C-0787(A), 097.C-0591(A), 1100.C-0481(H),
    0101.C-0350(A) and 2101.C-5048(A).}}

\author{D. Mesa\inst{1,2}, M. Bonnefoy\inst{3}, R. Gratton\inst{1}, G. Van Der Plas\inst{3}, V. D'Orazi\inst{1}, E. Sissa\inst{1}, A. Zurlo\inst{4,5,6}, E. Rigliaco\inst{1}, T. Schmidt\inst{7,8}, M. Langlois\inst{9,6},  A. Vigan\inst{6}, M.G. Ubeira Gabellini\inst{10}, S. Desidera\inst{1}, S. Antoniucci\inst{11}, M. Barbieri\inst{2}, M. Benisty\inst{12,3}, A. Boccaletti\inst{8}, R. Claudi\inst{1}, D. Fedele\inst{13}, D. Gasparri\inst{2}, T. Henning\inst{14}, M. Kasper\inst{15}, A.-M. Lagrange\inst{3}, C. Lazzoni\inst{1}, G. Lodato\inst{10}, A.-L. Maire\inst{14}, C.F. Manara\inst{15}, M. Meyer\inst{16}, M. Reggiani\inst{17}, M. Samland\inst{14}, M. Van den Ancker\inst{15}, G. Chauvin\inst{3}, A. Cheetham\inst{18}, M. Feldt\inst{14}, E. Hugot\inst{6}, M. Janson\inst{14,19}, R. Ligi\inst{6,20}, O. M\"oller-Nilsson\inst{14}, C. Petit\inst{21}, E.L. Rickman\inst{18}, F. Rigal\inst{22}, F. Wildi\inst{18}}

    \institute{\inst{1}INAF-Osservatorio Astronomico di Padova, Vicolo dell'Osservatorio 5, Padova, Italy, 35122-I \\
      \inst{2}INCT, Universidad De Atacama, calle Copayapu 485, Copiap\'{o}, Atacama, Chile\\
      \inst{3}Univ. Grenoble Alpes, CNRS, IPAG, 38000 Grenoble, France\\
      \inst{4}Nucleo de Astronomia, Facultad de Ingenieria y Ciencias, Universidad Diego Portales, Av. Ejercito 441, Santiago, Chile \\
      \inst{5}Escuela de Ingenieria Industrial, Facultad de Ingenieria y Ciencias, Universidad Diego Portales, Av. Ejercito 441, Santiago, Chile \\
      \inst{6}Aix Marseille Universit\'{e}, CNRS, LAM – Laboratoire d'Astrophysique de Marseille, UMR 7326, F-13388 Marseille, France \\
      \inst{7}Hamburger Sternwarte, Gojenbergsweg 112, 21029 Hamburg, Germany \\
      \inst{8}LESIA, Observatoire de Paris, PSL Research University, CNRS, Sorbonne Universit\'{e}s, UPMC Univ. Paris 06, Univ. Paris Diderot, Sorbonne, Paris Cit\'{e}, 5 Place Jules Janssen, 92195 Meudon, France \\
      \inst{9}Univ. Lyon, Univ. Lyon 1, ENS de Lyon, CNRS, CRAL UMR 5574, 69230 Saint-Genis-Laval, France \\
      \inst{10}Dipartimento di Fisica, Universit\'a Degli Studi di Milano, Via Celoria, 16, I-20133 Milano, Italy \\
      \inst{11}INAF-Osservatorio Astronomico di Roma, via di Frascati 33, 00078 Monte Porzio Catone, Italy \\
      \inst{12}Unidad Mixta Internacional Franco-Chilena de Astronomía (CNRS, UMI 3386), Departamento de Astronomía, Universidad de Chile, Camino El Observatorio 1515, Las Condes, Santiago, Chile \\
      \inst{13}INAF-Osservatorio Astrofisico di Arcetri, Largo E. Fermi 5, I-50125, Firenze, Italy \\
      \inst{14}Max-Planck-Institut f\"ur Astronomie, K\"onigstuhl 17, 69117, Heidelberg, Germany \\
      \inst{15}European Southern Observatory (ESO), Karl-Schwarzschild-Str. 2, 85748 Garching bei M\"unchen, Germany \\
      \inst{16}Department of Astronomy, University of Michigan, 1085 S. University Ave, Ann Arbor, MI 48109-1107, USA \\
      \inst{17}Space sciences, Technologies \& Astrophysics Research (STAR) Institute, Universit\'{e} de Liege, All\'{e}e du Six Aout 19c, B-4000 Sart Tilman, Belgium \\
      \inst{18}Geneva Observatory, University of Geneva, Chemin des Maillettes
51, 1290 Versoix, Switzerland \\
      \inst{19}AlbaNova University Center, Stockholm University, Stockholm, Sweden \\
      \inst{20}INAF-Osservatorio Astronomico di Brera, Via E. Bianchi 46, I-23807, Merate, Italy \\
      \inst{21}DOTA, ONERA, Universit\'e Paris Saclay, F-91123, Palaiseau France \\
      \inst{22}Anton Pannekoek Institute for Astronomy, Science Park 904, NL-1098 XH Amsterdam, The Netherlands \\
    }

   \date{Received  / accepted }

\abstract
   {}
   {R Coronae Australis (\rcra) is the brightest star of the Coronet
     nebula of the Corona Australis (CrA) star forming region. It has very red
     colors, probably due to dust absorption and it is strongly variable. High
     contrast instruments allow for an unprecedented direct exploration of the
     immediate circumstellar environment of this star.
   }
   {We observed \rcra\ with the near-IR channels (IFS and IRDIS) of SPHERE at
     VLT. In this paper, we used four different epochs, three of them from
     open time observations while one is from the SPHERE guaranteed time.
     The data were reduced using the DRH pipeline and the SPHERE Data Center.
     On the reduced data we implemented custom IDL routines with the aim to
     subtract the speckle halo. We have also obtained pupil-tracking H-band
     (1.45-1.85~\mic) observations with the VLT/SINFONI near-infrared
     medium-resolution (R$\sim$3000) spectrograph.
   }
   {A companion was found at a separation of 0.156\as from the star in the
     first epoch and increasing to 0.184\as in the final one.
     Furthermore, several extended structures were found around the star, the
     most noteworthy of which is a very bright jet-like structure North-East
     from the star. The astrometric measurements of the companion in the four
     epochs confirm that it is gravitationally bound to the star. The SPHERE
     photometry and the SINFONI spectrum, once corrected for extinction,
     point toward an early M spectral type object with a mass between 0.3 and
     0.55~\MSun. The astrometric analyis provides constraints on the
     orbit paramenters: e$\sim$0.4, semi-major axis at 27-28~au,
     inclination of $\sim70^{\circ}$ and a period larger than 30 years. We
     were also able to put constraints of few \MJup on the mass of
     possible other companions down to separations of few tens of au.
   }
{}


\titlerunning{Exploring \rcra\ with SPHERE}
\authorrunning{Mesa et al.}
   \maketitle
%

\section{Introduction}
\label{intro}
Nowadays young stars surrounded by a protoplanetary disk are considered as the
primary environment where studying the formation of giant exoplanets
\citep[see e.g.][]{2012ApJ...756..133C,2014A&A...565A..15M}.
In addition, it is clear that giant planets at large separations are more
frequently found around intermediate mass stars than around solar-mass stars
\citep[see e.g.][]{2010PASP..122..905J,2016PASP..128j2001B}.
Herbig AeBe (HAeBe) stars \citep{1960ApJS....4..337H,1992ApJ...397..613H}
are young ($<$10 Myr), intermediate mass (1.5-8\MSun) and typically with A,
B, F spectral types. They are characterized by a strong infrared excess
\citep{1994ASPC...62...23T} due to the presence of a circumstellar
protoplanetary disk. They are also experiencing accretion processes and their
spectral line profiles are very complex. These characteristics make
these objects very valuable in the field of extrasolar planets given that
they allow to explore the very first stages of the formation of planetary
systems. \par
The HAeBe star R Coronae Australis (\rcra\ - HIP\,93449) is the brighest
star in the very young, compact ($\sim$1 pc in diameter) and obscured Coronet
protostar cluster \citep{1984MNRAS.209P...5T}. This cluster is
characterized by a very high and spatially
variable extinction with $A_V$ up to 45 mag \citep{2008hsf2.book..735N}. This
nebula is located toward the center of the Corona Australis (CrA) molecular
clouds complex \citep{1992lmsf.book..185G}, one of the nearest star forming
region with a distance of $\sim$130~pc \citep{2008hsf2.book..735N}. More
  recently, \citet{2018ApJ...867..151D}, using the mean trigonometric
  parallax obtained from Gaia Data Release 2 \citep{2018yCat.1345....0G},
  estimated a distance of $154\pm4$~pc.
The presence of emission line objects was also signaled in the vicinity of
\rcra \citep{1993PASP..105..561G}. The CrA region was explored through deep
infrared imaging by \citet{1997AJ....114.2029W} that identified hundreds
of point-like sources. Moreover, they gave particular relevance to the study
of the region of the Coronet nebula where they identified extensive reflective
nebulae, dust-free cavities and Herbig-Haro objects. More recently,
\citet{2009PASP..121..350M} obtained the infrared spectra for a magnitude
limited sample of stars of the CrA region allowing their characterization. \par
The spectral type of \rcra\ is largely debated and varies from F5
\citep[e.g., ][]{1992ApJ...397..613H} and A5
\citep[e.g., ][]{1997ApJ...478..295C},
to B5III peculiar \citep{2006AJ....132..161G} and finally B8
\citep[e.g., ][]{1992A&A...260..293B}. Other physical charactistics of the star
are discussed in more detail in Sec.~\ref{s:star}. 
It is in a very early evolutionary phase \citep{1998A&A...331..211M} with an
age of $1^{+1}_{-0.5}$~Myr \citep{2011ApJ...736..137S} and still embedded in its
dust envelope whose emission dominates its SED from mid-infrared to millimeter
wavelengths \citep{2009A&A...508..787K}. The mass of the envelope was estimated
to be $\sim$10~\MSun by \citet{1993ApJ...406..674N} with an outer radius of
0.007~pc corresponding to $\sim$1450~au. \citet{2000MNRAS.319..337C} found a
high degree of linear ($\sim$8$\%$) and circular ($\sim$5$\%$) polarization
as a hint of scattering from dust grains. The polarization map in V band
also indicates the presence of an extended ($\sim$10\as corresponding to more
than 1500~au) disk-like structure
with a roughly north-south orientation \citep{1985MNRAS.215..537W}. A similar
orientation was found by \citet{2009A&A...508..787K} using interferometry in
the NIR with VLTI/AMBER that also found asymmetries in the brightness
distribution and obtained a disk inclination of $\sim$$35^{\circ}$. \par
Moreover, \citet{2003A&A...397..675T} proposed the presence of a stellar
companion and of an outflow based on their spectro-astrometric observations.
They estimated for the binary a separation of $\sim$8~au and a period of
$\sim$24~years. The presence of a stellar companion was also proposed by
\citet{2006A&A...446..155F} to explain the X-ray spectrum of \rcra\ based on
the presence of a strong X-ray emission that is not expected for HAeBe stars.
\par
In this paper we present the SPHERE \citep{2008SPIE.7014E..18B} view of \rcra\
and its closeby environment. The paper is organized as
follows: in Section~\ref{s:obs} we describe the observations and the data
reduction, in Section~\ref{s:star} we discuss the characteristics of \rcra,
in Section~\ref{s:res} we report our results that are then discussed
in Section~\ref{s:dis}. Finally, in Section~\ref{s:conclusion} we present our
conclusions.

\section{Observations and data reduction}
\label{s:obs}

\subsection{SPHERE data}
\label{spheredata}

\rcra\ was observed with SPHERE, the extreme adaptive optics instrument
operating at ESO very large telescope (VLT), in four epochs.
In Table~\ref{t:obs} we list all the SPHERE observations used in the present
analysis. All these observations were obtained in coronographic mode to be able
to suppress the stellar light. The star was observed in a first epoch in
the night of 2015-06-10
as part of the open time program 095.C-0787(A) (P.I. G. Van der Plas) aimed to
the analysis of disks of HAeBe stars. The observations were made in IRDIFS\_EXT
mode that is with IFS \citep{Cl08} operating between Y and H spectral bands
(0.95-1.65\mic) and IRDIS \citep{Do08} operating in dual band configuration
with the K1-K2 filters
\citep[K1=2.110\mic and K2=2.251\mic; ][]{2010MNRAS.407...71V}. Moreover, they
were taken in field-stabilized mode and in bad weather conditions
(see Table~\ref{t:obs} for more informations), limiting the contrast that can
be reached to a value of $\sim4\times10^{-4}$ at 0.4\as. \par A second epoch was
obtained during the night of 2016-08-10 for the open time program 097.C-0591(A)
(P.I. T Schmidt). In this case, the observations were made with IFS operating
in Y and J spectral bands (between 0.95 and 1.35\mic) and IRDIS operating with
the H broad band filter (central wavelength 1.625\mic with a width of
0.290\mic) instead of the standard H2-H3 dual band filter. This
was a short observation with a total field rotation of $24.7^{\circ}$ and it was
taken in weather conditions comparable to those from the first observing epoch.
\par
A third epoch was obtained on 2018-06-19 in the framework of the SHINE
\citep[SpHere INfrared survey for Exoplanets; ][]{2017sf2a.conf..331C}
guaranteed time observations (GTO). Like for the first epoch, these
observations were made in IRDIFS\_EXT mode. In this case
the weather condition were much better than in the previous epochs
(Table~\ref{t:obs}) and the observations were performed in pupil stabilized
mode with a total rotation of $71.7^{\circ}$ allowing the implementation of high
contrast imaging methods like angular differential imaging
\citep[ADI; ][]{2006ApJ...641..556M} and spectral differential imaging
\citep[SDI; ][]{1999PASP..111..587R}. This allowed to reach a much better
contrast of $6.7\times10^{-6}$ at 0.4\as. \par
Finally, we obtained data in a fourth
epoch from the open time program 0101.C-0350(A) (P.I. M.G. Ubeira Gabellini).
Also in this case the observations were done with SPHERE operating in
IRDIFS\_EXT mode and in conditions very similar to the third epoch. The total
rotation was in this case of $\sim90^{\circ}$.
For all the epochs, in addition to the coronagraphic observations, we also
performed observations with satellite spots symmetric with respect to the
central star with the aim to better define the position of the star behind the
coronagraph using the method first proposed by \citet{2006ApJ...647..620S} and
\cite{2006ApJ...647..612M}. The use of these spots in the context of the SPHERE
observations is discussed in \citet{2013aoel.confE..63L} and in
\citet{mesa2015}. Furthermore, observations of the star outside the coronagraph
were taken to perform the photometric calibration of the observations. These
were observed with an appropriate neutral density filter to avoid saturation of
the PSF. However, this type of calibration images was not taken in the
case of the 2016-08-10 observations, therefore it was not possible to define
the contrast of the image and they were only used for astrometric measures
using a synthetic Gaussian PSF created for this purpose. \par
The GTO observations were reduced using the SPHERE data center
\citep{2017sf2a.conf..347D} applying the appropriate calibrations following
the data reduction and handling \citep[DRH; ][]{2008SPIE.7019E..39P} pipeline.
For the open time observations, instead, we applied directly the DRH pipeline
and IDL custom routines for IFS without the intermediation of the
SPHERE data center. The required calibration
in the case of IRDIS are aimed to the creation of the master dark and of the
master flat-field frames and to the definition of the star center. The steps
for the IFS data reduction, instead, include the dark and flat-field
correction, the definition of the position of each spectra on the detector,
the wavelength calibration and the application of the instrumental flat. On
the reduced data, we then performed algorithms like principal components
analysis \citep[PCA; ][]{2012ApJ...755L..28S} and template locally optimized
combination of images \citep[TLOCI; ][]{2014SPIE.9148E..0UM} with the aim to
implement speckle subtraction methods like ADI and SDI. To this aim we used
both the IDL procedures described in \citet{mesa2015} and \citet{zurlo2014}
and the consortium pipeline application called SpeCal
\citep{2018A&A...615A..92G}.

\begin{table*}[!htp]
  \caption{List and main characteristics of the SPHERE observations of \rcra\
    used for this work.}\label{t:obs}
\centering
\begin{tabular}{ccccccccc}
\hline\hline
Date  &  Obs. mode  & field/pupil  & DIMM seeing & $\tau_0$ & wind speed & Rotation & DIT & Total Exposure\\
\hline
2015-06-10  & IRDIFS\_EXT &  field & 1.35\as & 2.2 ms & 10.2 m/s & none         & 16 s & 1680 s\\
2016-08-10  & YJ+BB\_H    &  pupil & 1.37\as & 2.9 ms & 12.5 m/s &$24.7^{\circ}$ & 64 s & 1920 s \\
2018-06-19  & IRDIFS\_EXT &  pupil & 0.55\as & 6.5 ms &  7.1 m/s &$71.7^{\circ}$ & 96 s & 4608 s \\
2018-08-16  & IRDIFS\_EXT &  pupil & 0.76\as & 4.6 ms &  7.8 m/s & $90.5^{\circ}$ & 12 s & 5760 s \\
\hline
\end{tabular}
\end{table*}

\subsection{SINFONI data}
\label{sinfonidata}

In this work we also made use of observations of the \rcra\ system obtained
with the AO-fed integral field spectrograph SINFONI
\citep{2003SPIE.4841.1548E,2004SPIE.5490..130B} during the night of 2018-09-11
(2101.C-5048(A); P.I.: D. Mesa). The spatial sampling was 0.0125\as/pixel
$\times$ 0.025\as/pixel for a total FOV of 0.8\as$\times$0.8\as. We observed
the target with H-band grating operating between 1.45 and 1.85~\mic with a
resolution R$\sim$3000. Given the brightness of the star, to avoid saturation we
adopted the minimum DIT of 0.83 s. We took 30 datacubes each composed by
40 frames for a total exposure time of 996 s. Moreover, to be able to implement
high-contrast imaging techniques like ADI, we observed the target using the
pupil tracking mode for a total rotation of the FOV of $\sim95^{\circ}$. \par
The data were reduced using the version 3.1.1 of the SINFONI pipeline
\footnote{\url{http://www.eso.org/sci/software/pipelines/sinfoni/sinfoni-pipe-recipes.html}}. The science data were corrected for bad and non-linear
pixels and for distortion. The final calibrated datacubes were then
reconstructed from the associated wavelength calibration. On the final
calibrated datacubes we then corrected the wavelength dependent drift due to
the atmospheric refraction following the same procedure described in
\citet{2015MNRAS.453.2378M} and \citet{2018A&A...617A.144H}. We then
subtracted the stellar halo at each of the 2120 wavelengths applying both
the classical ADI and PCA techniques. For the SINFONI data, we did not use
the SDI technique because, while it could help in gaining in S/N, on the other
hand it could introduce deformations in the object spectrum. Given that
the main aim of the SINFONI data is a better classification of the companion
object through its spectral lines so that the use of the SDI could lead to a
biased result. To derive the correct parallactic
angle to be associated to each datacube of our dataset we took the mean of
the values of the parallactic angle at the beginning and at the end of each
exposure that are given in the header associated with the datacube. These values
were then corrected by the value of $304^{\circ}$ to properly orientate them
with North up and East to the right (see \citealt{2015MNRAS.453.2378M} for the
details of view orientation of SINFONI pupil-tracking mode). 

\section{Stellar parameters}
\label{s:star}
In this Section we discuss the stellar parameters that are important for
our analysis like e.g. the distance, the luminosity and the variability.
The parallax of the star as measured by Gaia in Data Release 2
\citep{2018yCat.1345....0G} is 10.53$\pm$0.70~mas corresponding to a
distance of $94.9^{+6.7}_{-5.9}$~pc that is far from the previous estimated
distance of the CrA region \citep[d$\sim$130~pc; ][]{1999AJ....117..354D}.
Then, when looking for other objects in the CrA region in the Gaia archive we
have found lower values for the parallax. To perform this research we have
selected into the Gaia archive objects with angular separation lower than 10
arcmin from \rcra and a value of the parallax larger than 5~mas. Moreover, we
have selected all the stars with a proper motion in RA smaller than 5~mas/yr
and a proper motion in Declination smaller than -25~mas/yr with the aim to
retain only stars with a proper motion similar to that of \rcra. In this way
we have selected 15 objects and we have verified, through kinematics and
literature searches, for each of them that it was
part of the CrA region. The median value of the parallax for these objects is
$6.54\pm0.33$~mas corresponding to a distance of $152.9^{+8.1}_{7.3}$~pc.
Given that the membership of \rcra\ to the CrA region is well established, the
Gaia value for \rcra\ is probably plagued by an error possibly due to the fact
that the star is deeply embedded in its dust envelope. For this reason we
decided to assume, for the rest of this work, the distance obtained from the
median of the other objects of the CrA region. It is worth to note that
the adopted distance for \rcra\ is very similar to that of the Corona
Australis region derived by \citet{2018ApJ...867..151D}.\par
Another important characteristic of this star is its higher luminosity in the
NIR \citep[J=6.94; H=4.95; K=3.46;][]{2002yCat.2237....0D,2003yCat.2246....0C}
than at optical wavelengths \citep[V=11.92;][]{2010MNRAS.403.1949K}. However it
is also known that this star is strongly variable. So far the only published
homogeneous photometric survey of \rcra\ was carried out at the visible
wavelengths exploiting the Maidanak Observatory in Uzbekistan starting from
1983, and was published by \citet{1999AJ....118.1043H}. An analysis of these
data using a generalized Lomb-Scargle periodogram
\citep[GLSP- ][]{2009A&A...496..577Z} shows several significative peaks, with
separations in frequency corresponding to one year (0.00276 dex). The two
highest power frequency peaks are those of 55.7 days and 65.9 days, and
cannot be due to the data sampling.  Furthermore,
\citet{2010PASP..122..753P} studied the photometric stability of \rcra\
exploiting a 100-year long visual data sequence provided by the American
association of variable star observers (AAVSO). They found a period of 66 days,
stable in time but slightly variable in amplitude. The amplitude of this
variability was larger than 4 magnitudes at visible wavelengths.
\par
Finally, a fundamental parameter to be analyzed is the extinction in the
direction of \rcra\ that can strongly affect the extracted spectrum of the
star and of its companion (see Section \ref{s:res}). \citet{1992A&A...260..293B}
found a value of E(B-V)=0.99 and a value of $R_V$=4.7 corresponding to an
absorption of $A_V$=4.65. More recently, \citet{2006A&A...459..837G} found
a lower value of $A_V$=1.40 while \citet{2017A&A...599A..85L} found a value
of $A_V$=3.31 assuming a value of 3 for $R_V$. However, these values
were determined for a wide region
around the star while the extinction is strongly variable in the Coronet
region. For this reason we developed a method to estimate the
extinction in the direction of the companion that will be described in
Section~\ref{s:charcom}.


\section{Results}
\label{s:res}

\subsection{Companion}
\label{s:companion}

\begin{figure*}[!htp]
\centering
\includegraphics[width=0.45\textwidth]{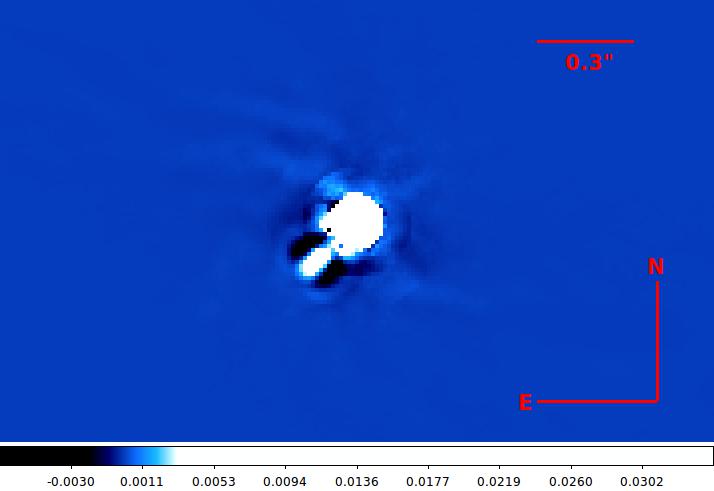}
\includegraphics[width=0.45\textwidth]{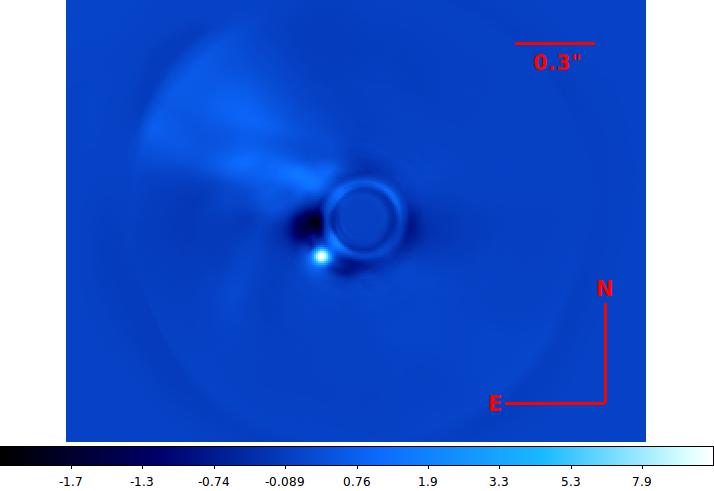}
\caption{Final images for \rcra\ system obtained from the best SPHERE dataset
  taken on 2018-06-19 (chosen because it is the best quality observation epoch).
  Both of them are obtained from a median of the images at all the
  wavelengths covered by the two instruments. {\it Left:} Final image obtained
  with IRDIS. {\it Right:} Final images obtained with IFS. Both images have
  flux scales in such a way to make the presence of the companion clear. In
  both images there is however hints of the presence of the jet-like structure
  described in Section~\ref{f:doublejet}.}
\label{f:finalim_comp}
\end{figure*}

A companion is found in the final images of all the epochs of observations.
In Figure~\ref{f:finalim_comp} we show the final images that we obtained from
the 2018-06-19 data that, as explained in Section~\ref{s:obs}, allowed to
obtain much deeper images. The scales used in these images have been chosen
to better visualize the companion but it is however apparent the presence
of extended structures that we will discuss much in detail in
Section~\ref{s:jet}. The companion is less visible in the data taken on
2016-08-10, maybe due to the shorter wavelength set up chosen for this
observation coupled with a very red spectrum of the source as we will show
in Section~\ref{s:companion}. \par
\begin{figure}
\centering
\includegraphics[width=\columnwidth]{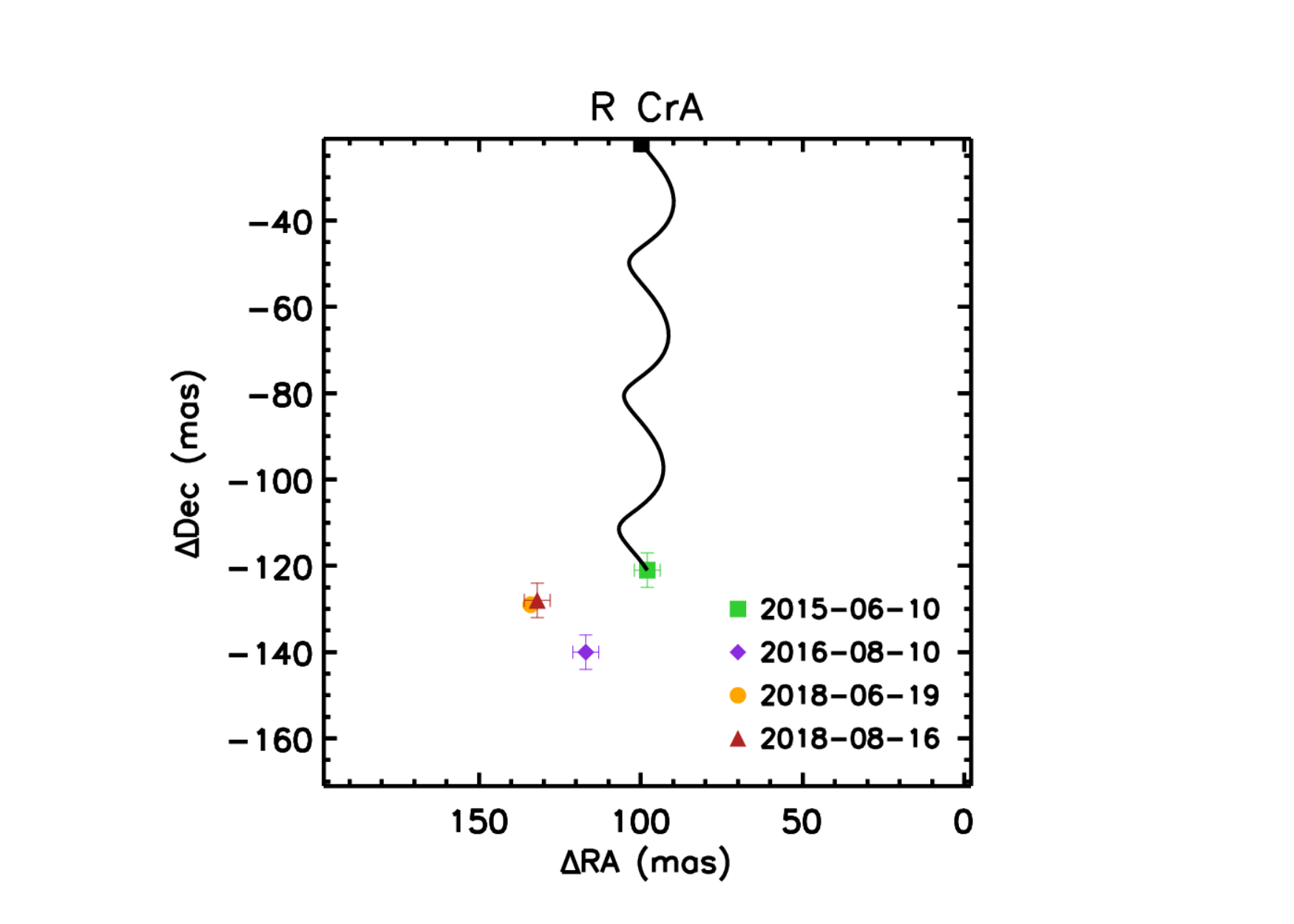}
\caption{Relative astrometric position of \rcra\,B with respect to the host
  star at four different epochs. In black it is shown the path the object
  would have followed in the three years between the first and the fourth
  epochs. The black square corresponds to the expected position for the
  companion at the epoch of the third observation if it were a stationary
  background object.}
\label{f:astro}
\end{figure}

In Table~\ref{t:astro} we list the relative astrometric values for the companion
while in Figure~\ref{f:astro} we display the relative positions of the
companion with respect to the star, compared to the position that it would
have had if it were a stationary background object, demonstrating that
it is actually comoving with the star. We interpret the variations in the
companion separation and position angle as evidences of its orbital motion.
To further confirm this, we have to consider that the scatter in velocity
for CrA members is of the order of 1-2~km/s while the scatter in
velocity between the companion and the central star is between 5 and 10~km/s.
This is an other hint that we are actually observing the orbital motion of the
companion around the central star.\par Two main problems affect the
extraction of the photometry. Both of them are
linked to the characteristics of the star itself. The first one
is due to the large difference in magnitude according to the considered spectral
band. Given the high brightness of the star in the H and K
band, we had to use the strongest neutral density filter (ND3.5) for the
observation of the off-axis PSF to avoid saturation of the star (see
Section~\ref{s:obs}). In this way, however, the star was no longer visible in
the IFS Y and J band channels. This of course does not allow a proper
flux calibration at these wavelengths. Moreover, the star is strongly
variable and for this reason it is difficult to define the absolute magnitude
of the companion starting from the contrast (when possible to retrieve). To
overcome these problems and to obtain a reliable spectrum of the companion, we
have then compared the PSF of \rcra\ at each
wavelength obtained from the observation of 2018-06-19, when available, to the
PSFs of other stars taken during the same night in similar conditions. They are
HIP\,63847 and HIP\,70833 that are not known as variable stars. The results are
similar for both stars and they allow to define a value for H of
$6.95\pm0.29$~mag and for K of $4.63\pm0.32$~mag for \rcra\ at the epoch of our
observation. Also, it was possible to put upper limits to the Y and J
magnitudes: Y$>$10.92~mag and J$>$8.36~mag.
The comparison of these
values with those given in literature (see Section~\ref{s:star}) put our
observation at (or very near to) the minimum of the star brightness variability.
Given the comparable results obtained previously we used as reference the PSF
of only HIP\,63847 to be able to extract
the spectrum in flux for \rcra\,B shown in Figure~\ref{f:spectrum}.
As a first step we calculated the spectrum in contrast with respect to the
reference star introducing a negative simulated planet at the companion
position and run our speckle subtraction procedure adjusting its flux in such
a way to minimize the standard deviation in a small region around the
companion position. We apply this procedure both using an ADI and a PCA-based
data reduction and we obtained comparable results. The contrast
spectrum was then converted to flux by multiplying it by a flux-calibrated
BT-NEXTGEN \citep{2012RSPTA.370.2765A} synthetic spectrum for HIP\,63847
adopting $T_{eff}$=5300~K, $\log{g}$=-0.5 and [M/H]=-0.5 that gives the best
fit to the SED of the star according to the VOSA tool
\citep{2008A&A...492..277B}.

\begin{table*}[!htp]
  \caption{Astrometric results obtained for \rcra\,B.}\label{t:astro}
\centering
\begin{tabular}{ccccc}
\hline\hline
Date  &  $\Delta$RA (\as)  & $\Delta$Dec (\as)  & $\rho$ (\as) & PA \\
\hline
2015-06-10  & 0.098$\pm$0.004 & -0.121$\pm$0.004 & 0.156$\pm$0.004 & 141.0$\pm$0.2 \\
2016-08-10  & 0.117$\pm$0.004 & -0.140$\pm$0.004 & 0.182$\pm$0.004 & 140.0$\pm$0.2 \\
2018-06-19  & 0.134$\pm$0.001 & -0.129$\pm$0.001 & 0.184$\pm$0.001 & 134.2$\pm$0.2 \\
2018-08-16  & 0.132$\pm$0.004 & -0.128$\pm$0.004 & 0.184$\pm$0.004 & 134.0$\pm$0.2 \\
\hline
\end{tabular}
\end{table*}

\subsection{Jet-like structure and circumstellar environment}
\label{s:jet}

\begin{figure*}[!htp]
\centering
\includegraphics[width=0.45\textwidth]{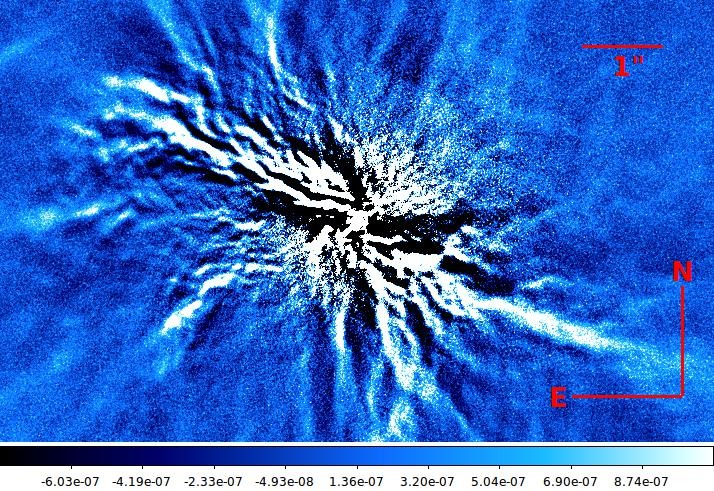}
\includegraphics[width=0.45\textwidth]{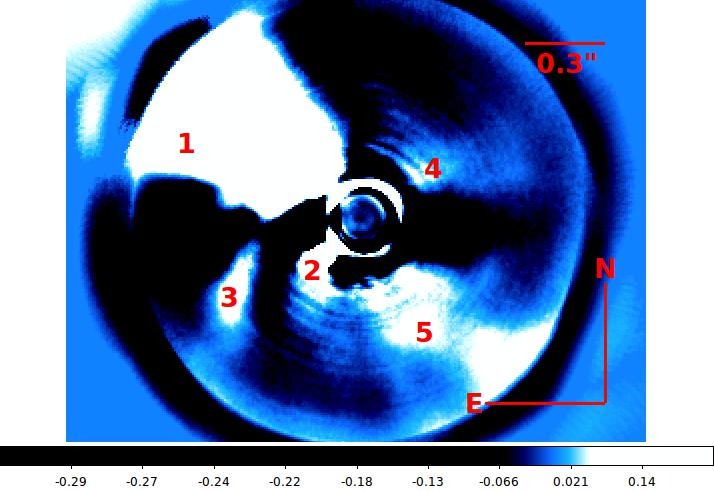}
\caption{Final images obtained for IRDIS using SpeCal TLOCI ({\it Left}) and
  IFS using PCA with 10 principal components ({\it Right}) scaled in such a
  way to enlight the presence of extended structures. In the IFS image we tag
  with 1 a jet-like structure, with 2 the position
  of the companion, with 3 and 4 two faint structures that could be part of
  the disk and with 5 a fainter counter-jet.}
\label{f:finalim_jet}
\end{figure*}

\begin{figure}
\centering
\includegraphics[width=\columnwidth]{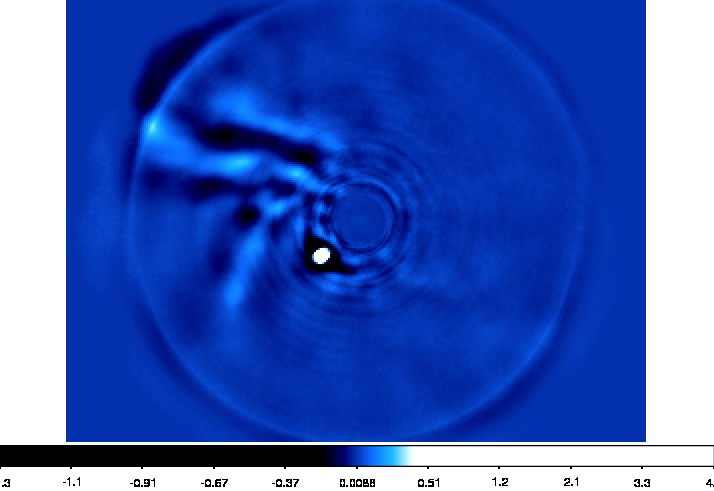}
\caption{Final IFS image after the subtraction of PCA 100 principal
  components aimed to highlight the presence of different substructure into
  the jet-like structure North-East from \rcra. Scale and orientation are the
  same as those on the right panel of Figure~\ref{f:finalim_jet}.}
\label{f:doublejet}
\end{figure}

The environment around \rcra\ is very rich of different extended
structures as it is shown in Figure~\ref{f:finalim_jet} where we display the
final images obtained with IRDIS and IFS with a scale thought to highlight the
presence of extended structures. The most striking of them is the jet-like
structure North-East from the star that is tagged with 1 in the IFS image
(right panel of Figure~\ref{f:finalim_jet}). The same structure is also clearly
visible in the IRDIS image (left panel of Figure~\ref{f:finalim_jet}) up to a
separation of 3.5\as corresponding to a projected separation of more than
535~au. The position angle of median axis of the jet-like structure, calculated
using the IFS image, is of $57.8\pm8.4^{\circ}$. The IRDIS image allows to
define an aperture of $\pm14^{\circ}$ at the maximum separation at which it is
visible. However, from the IFS image obtained subtracting 100 principal
PCA components displayed in Figure~\ref{f:doublejet}, it is clear that the
jet-like structure consists of at least two separated structures with position
angles of $66.1^{\circ}$ and
$49.4^{\circ}$ respectively. This is not an artefact of the differential imaging
because the two structures are clearly visible also on images where
no differential imaging was applied. The structure and the nature of this
jet-like structure will be studied in more detail in a dedicated paper
(Rigliaco et al., in prep.). A fainter counter-jet-like structure is also
visible both in the IRDIS and the IFS images at an angle of about $180^{\circ}$
with respect to the jet-like structure described above and it is tagged with 5
in the IFS image in Figure~\ref{f:finalim_jet}. Another
curved structure is visible South-East of the star and apparently it starts
from the base of the jet-like structure and not from the star itself. It is
tagged with 3 in
the IFS image in Figure~\ref{f:finalim_jet} but it is even more clearly visible
in Figure~\ref{f:doublejet}. Its approximated position angle is $103.5^{\circ}$
and it could be part of an external disk of which could also be part the very
faint structure tagged 4 North-West to the star with a position angle of
$305.9^{\circ}\pm9.0^{\circ}$ at $\sim180^{\circ}$ from the position of the
companion (tagged 2 in Figure~\ref{f:finalim_jet}).

\subsection{Results from SINFONI data}
\label{s:sinfores}

\begin{figure}
\centering
\includegraphics[width=\columnwidth]{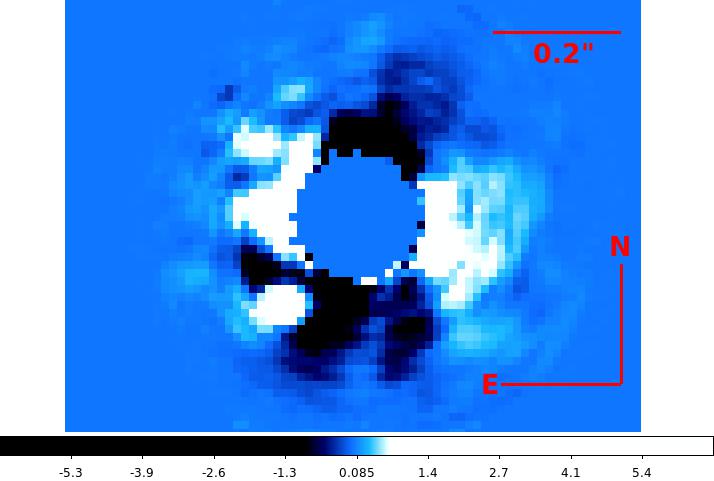}
\caption{\rcra\ image resulting from the median of the final SINFONI datacube
  obtained applying the ADI technique.}
\label{f:sinfoniimage}
\end{figure}

The companion was retrieved also in the SINFONI data as shown in the median
image of the final datacube obtained applying the ADI method that is
displayed in Figure~\ref{f:sinfoniimage}. From the final datacube, applying
aperture photometry at each single wavelength image, we were then able to
extract the spectrum for the companion that was then corrected for telluric
absorption using data from a B2 standard star (HD\,203617) observed during the
same night and at a similar airmass. The extracted SINFONI spectrum is very
noisy at shorter and longer wavelengths and, for this reason, the reduction of
the telluric absorption was not effective at those wavelengths and we decided
to retain only wavelengths between 1.50 and 1.75~\mic remaining with 1282
wavelengths that were then used to perform the cross-correlation procedure
described in Section~\ref{s:sinfonichar}.

\section{Discussion}
\label{s:dis}

\subsection{Characterization of the companion through evolutionary models}
\label{s:charcom}

\begin{figure}
\centering
\includegraphics[width=\columnwidth]{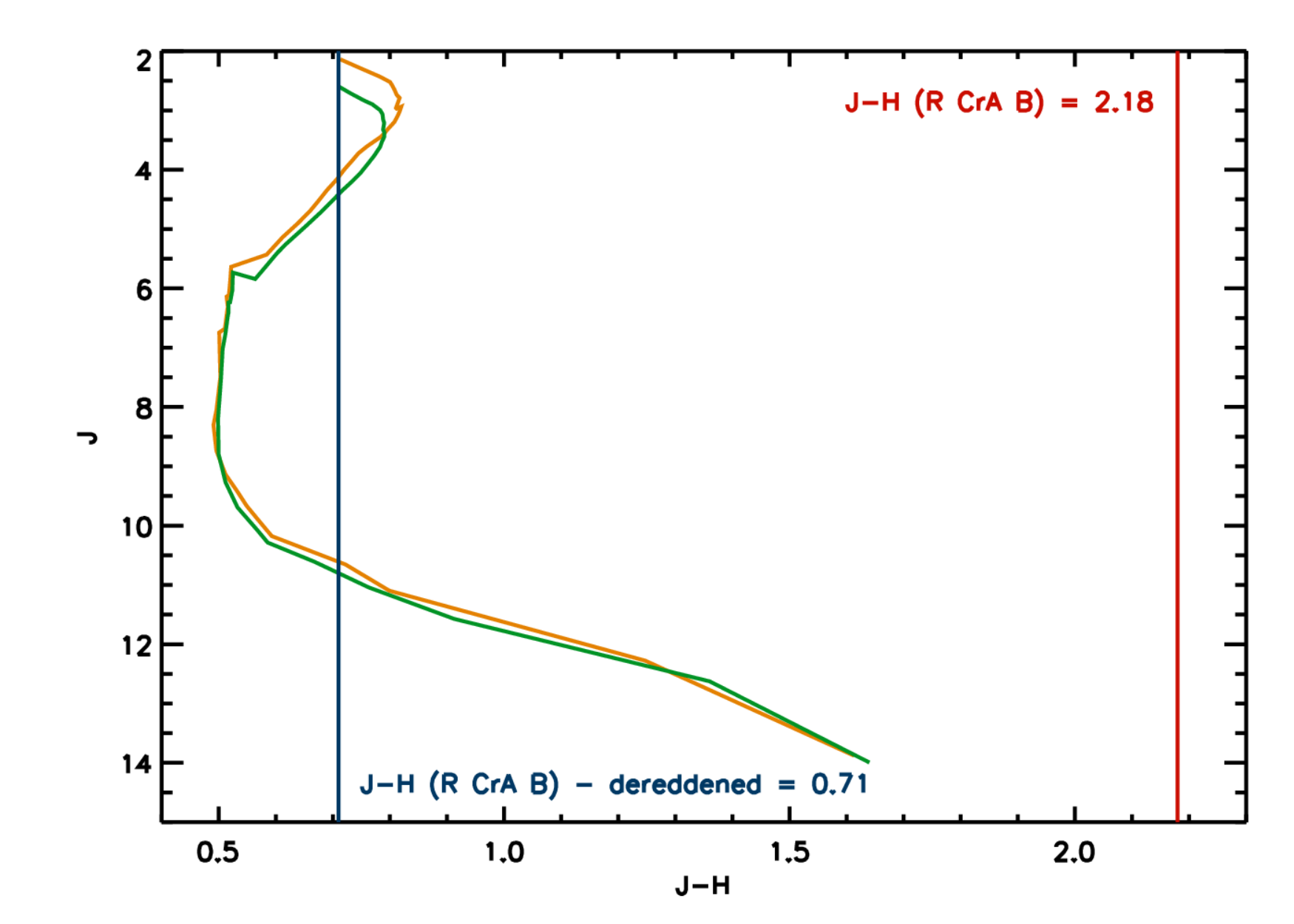}
\caption{Isochrones J vs J-H in the case of 1~Myr (orange line) and 2~Myr
  (green line). The red vertical line represents the J-H value calculated for
  \rcra\,B.}
\label{f:isochrone}
\end{figure}

The very strong and variable absorption of the CrA region does not allow
to blindly use the determinations of the absorption listed in
Section~\ref{s:star}. To overcome this problem we have exploited the
isochrones of the BT-Settl models \citep{2014IAUS..299..271A}. In
Figure~\ref{f:isochrone} we show the 1 and 2~Myr isochrones for J vs. J-H
compared with the value of J-H=$2.18\pm0.11$ (red vertical line) calculated
for the companion of \rcra\ from the extracted apparent magnitudes that are
listed in the second column of Table~\ref{t:photo}. As a first step we
calculated the value of the reddening E(J-H) taking as reference the median
value of J-H for the 1~Myr isochrone
in the region with 6$<$J$<$9 where the value of J-H is almost constant.
We then calculated the corresponding value of $A_J$ using the formula
from \citet{1989ApJ...345..245C}. With this value and applying the
appropriate correction for the distance modulus we can retrieve the absolute
magnitude in the J band that, through the use of the BT-Settl models, allow
to estimate the mass of the companion. Using the BT-Settl models we can
then obtain the absolute magnitude also for the H band and from this
calculate an updated value for J-H and for E(J-H) comparing the latter to
the original value for the companion. We then checked if the initial
value of E(J-H) corresponds to the new value that we obtain from the
model. We iterated the steps described above until the difference between the
input and the output values for E(J-H) is less than 0.01 mag. Once we
have obtained the final values for $A_J$ and for the absolute magnitude
in the J band, we  retrieved the same values also for all the other spectral
bands. These results are listed in column four and five of Table~\ref{t:photo}.
\par
Moreover, the absolute magnitudes listed in Table~\ref{t:photo} allow to
define some of the main physical
characteristics of \rcra\,B through the use of the BT-Settl models. We
obtained for the mass a value of $0.29\pm0.08$~\MSun, $T_{eff}$=$3270\pm175$~K,
$\log{g}$=$3.45\pm0.06$ and finally a radius of $16.7\pm4.2$~\RJup.
To evaluate the errors in these results we took into account the
uncertainties on the distance, on the age and on the magnitudes of the
host star.
According to \citet{2013ApJS..208....9P} the value of $T_{eff}$ that we found
would correspond to a spectral type of M3-M3.5. The relatively low surface
gravity found for the companion, coupled to the large radius, are hints for a
not fully completed gravitational collapse as expected
for such young objects. \par
The value of $A_J$ was used to derive an estimate of $A_V$=15.89 and, assuming
a value of $R_V$=4.7, we obtained a E(B-V)=3.38. These values were then used
to correct the spectrum obtained with the procedure described in
Section~\ref{s:companion} using the IDL ccm\_unred routine based on the
\citet{1989ApJ...345..245C} formula. While the uncorrected spectrum is very
red, after the correction we remain with a much bluer spectrum typical of a
stellar object. 

\begin{table*}[!htp]
  \caption{Photometric results obtained for \rcra\,B. The second column
    lists its apparent magnitudes before the correction for the extinction,
    the third column the differences in magnitudes with respect to the central
    star, the fourth column lists the values of the estimated extinction and
    finally the fifth column presents its absolute magnitudes after the
    correction for the extinction.}\label{t:photo}
\centering
\begin{tabular}{ccccc}
\hline\hline
Spectral band  & App. mag & $\Delta$Mag & $A_{\lambda}$ & Dered abs. mag.  \\
\hline
 Y   &  $ 16.45\pm0.05$ & $5.53\pm0.19$ & $5.83\pm0.32$ & $4.70\pm0.27$ \\
 J   &  $ 14.60\pm0.08$ & $6.24\pm0.74$ & $4.48\pm0.26$ & $4.20\pm0.26$ \\
 H   &  $ 12.42\pm0.03$ & $5.47\pm0.32$ & $3.01\pm0.38$ & $3.49\pm0.35$ \\
 K1  &  $ 10.63\pm0.01$ & $6.00\pm0.33$ & $1.36\pm0.30$ & $3.35\pm0.29$ \\
 K2  &  $ 10.27\pm0.01$ & $5.64\pm0.33$ & $1.18\pm0.28$ & $3.17\pm0.27$ \\
\hline
\end{tabular}
\end{table*}

\begin{figure}
\centering
\includegraphics[width=\columnwidth]{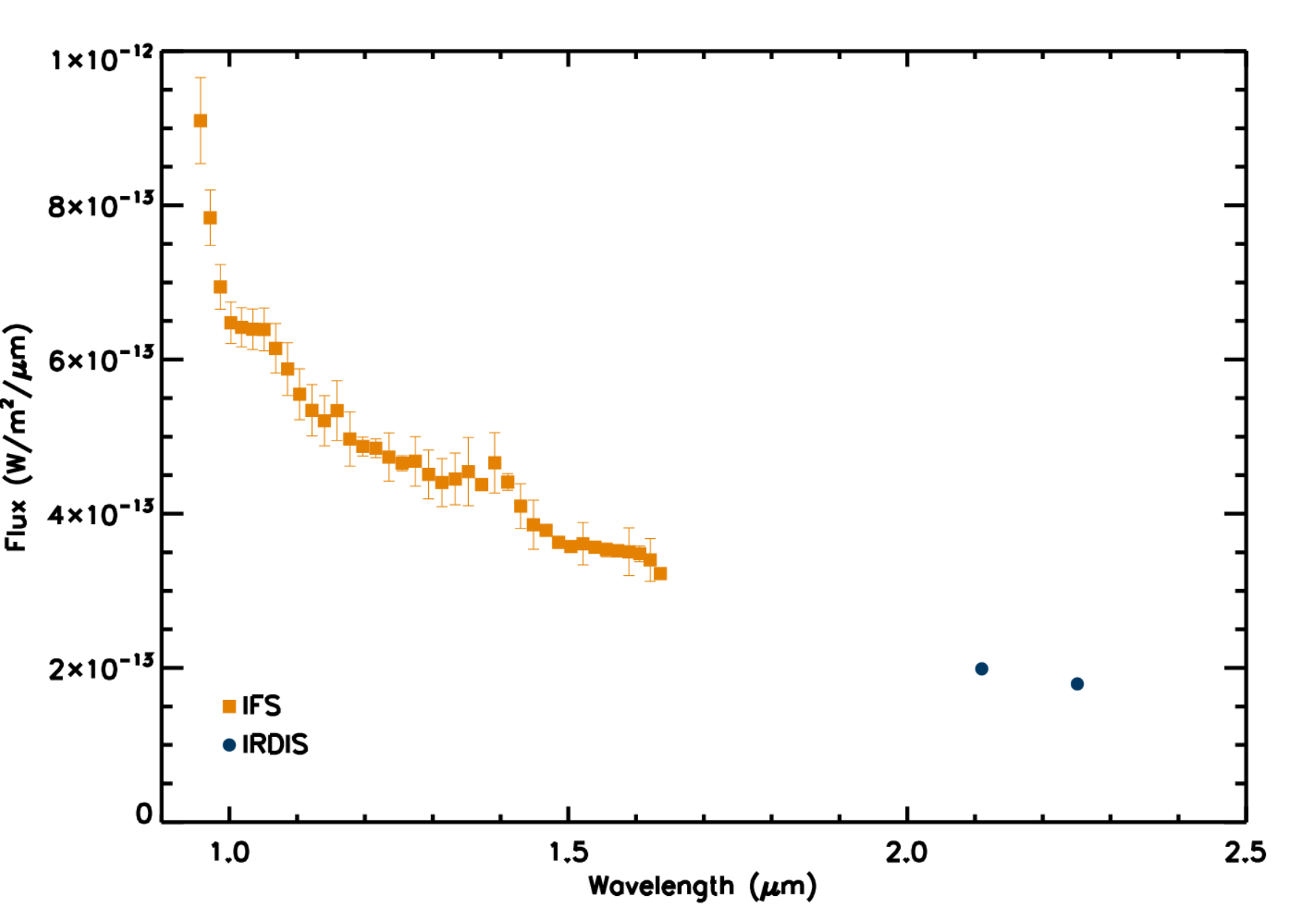}
\caption{Final spectrum extracted for \rcra\,B from SPHERE data after
  applying the correction for the extinction. The orange squares are measures
  obtained with IFS while the blue circles are obtained with IRDIS. In this
  last case the error bars are smaller than the symbol size.}
\label{f:spectrum}
\end{figure}

\subsection{Fitting with template spectra and atmospheric models}
\label{s:fitting}

\begin{figure}
\centering
\includegraphics[width=\columnwidth]{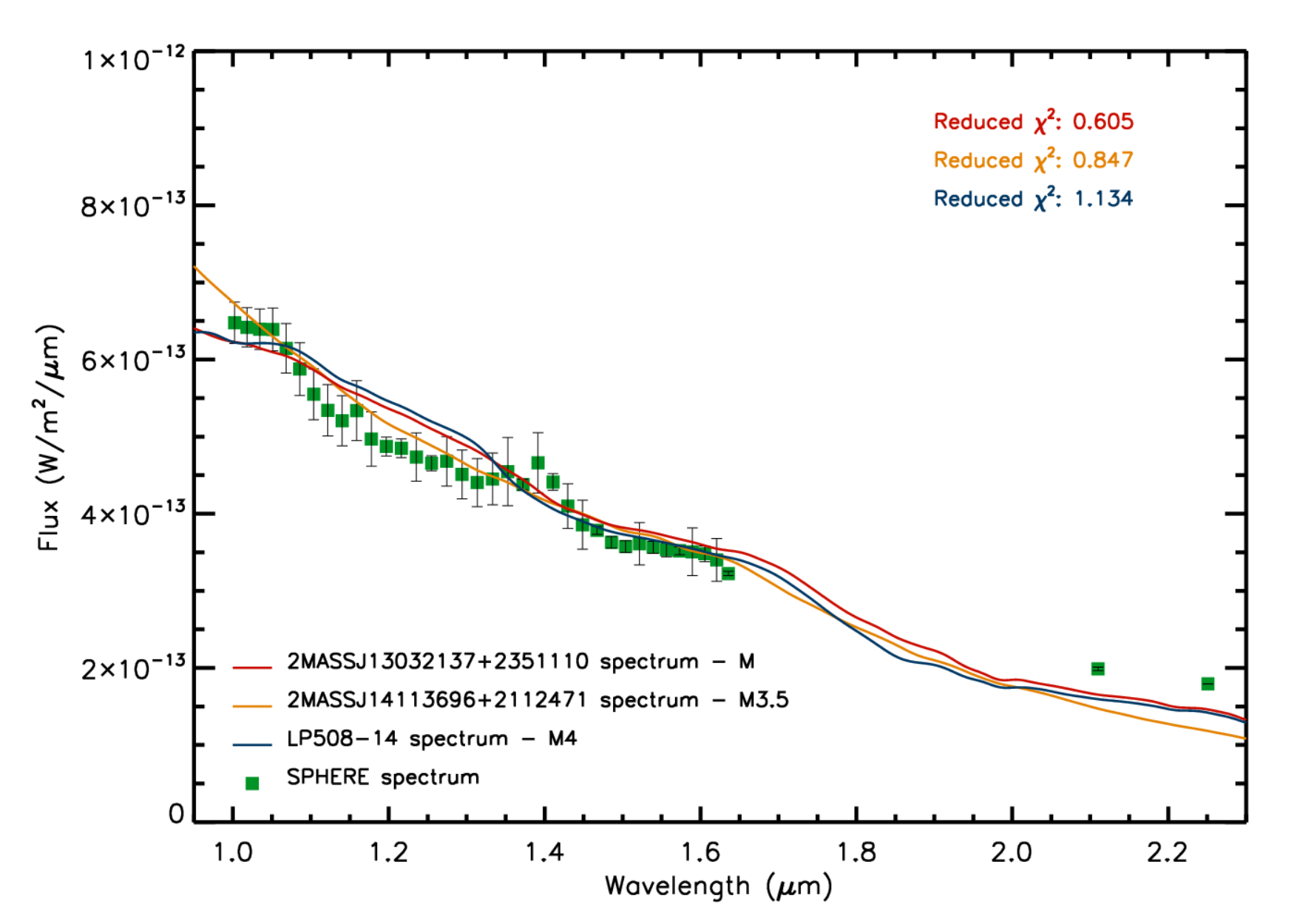}
\caption{Comparison of the \rcra\,B spectrum obtained from SPHERE (green
  squares) with three of the best fit template spectra.}
\label{f:template}
\end{figure}

To further characterize the companion we have tried to fit its extracted
spectrum with libraries of template spectra with the aim to define its
spectral type. To this aim we have used the library of field BDs spectra taken
from the {\it Spex Prism spectral Libraries}\footnote{\url{http://pono.ucsd.edu/~adam/browndwarfs/spexprism/}} \citep{2014ASInC..11....7B} and the
library of spectra taken from \citet{2013ApJ...772...79A}. The procedure
is similar to those adopted in \citet{2016A&A...593A.119M},
\citet{2018A&A...612A..92M} and \citet{2018AJ....156..182L}. For the fit
we did not use the three spectral points at shorter wavelengths that are
anomalously high as can be seen in Figure~\ref{f:spectrum}. This is
probably due to a contamination arising from the particular configuration
of the IFS raw data \citep[see e.g. ][]{Cl08}. In these data the faint
blue ends of the \rcra\,B spectrum in one spaxel are adjacent to
the very bright red ends of the spectrum on a near spaxel. This results in
some light from the bright red pixel leaking toward the faint blue pixels.
The final results for the fit procedure are shown in Figure~\ref{f:template}
and are compatible with an early-M spectral type for the companion. \par
\begin{figure}
\centering
\includegraphics[width=\columnwidth]{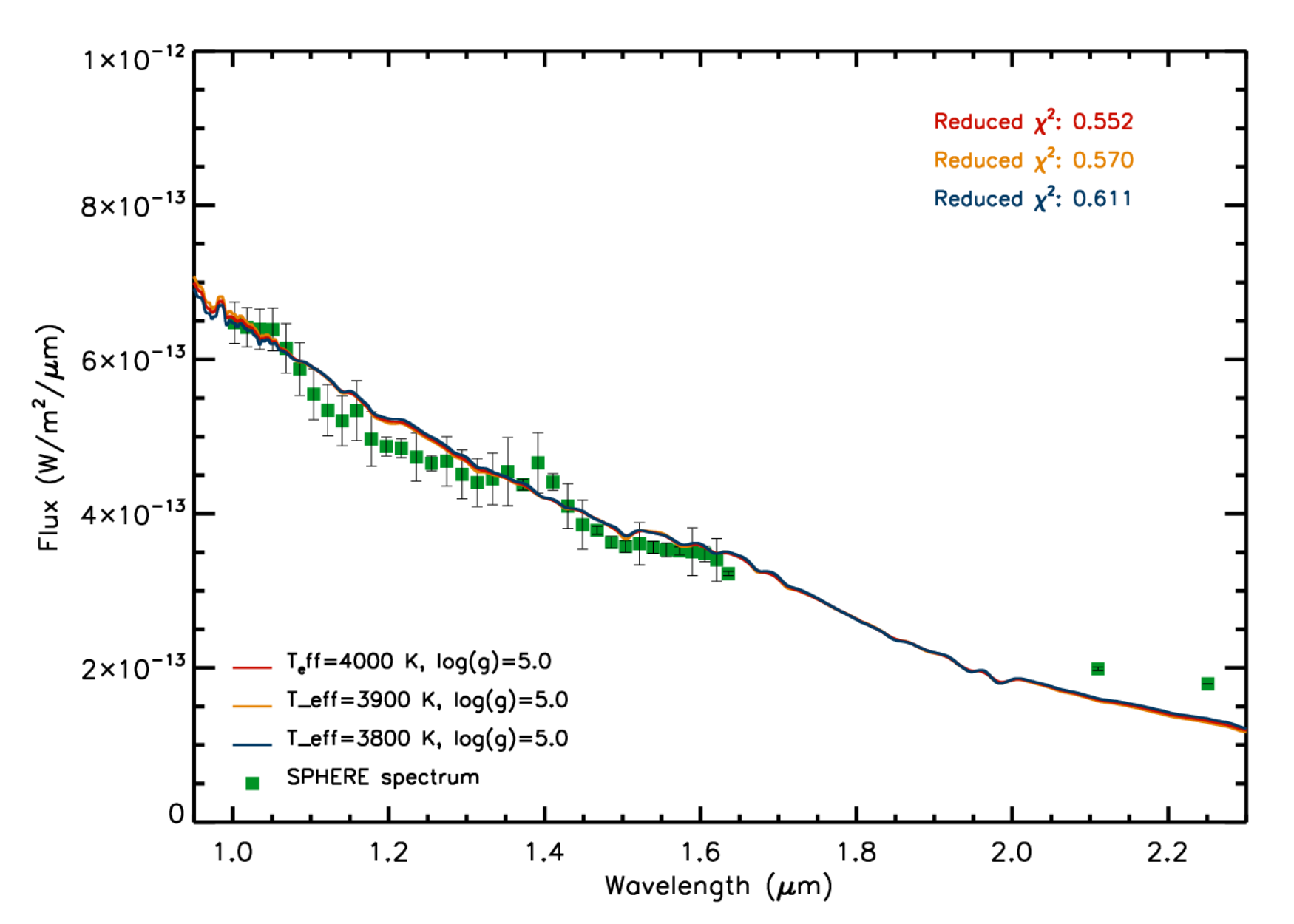}
\caption{Comparison of the \rcra\,B spectrum obtained from SPHERE (green
  squares) with three of the best fit BT-Settl atmospheric models.}
\label{f:synthmodel}
\end{figure}

We have also tried to fit the extracted spectra with a set of BT-Settl
and BT-NextGen atmospheric models
\citep{1997ARA&A..35..137A,2014IAUS..299..271A} with a grid covering $T_{eff}$
between 900 and 4000 K with a step of 100 K and a $\log{g}$ ranging between
2.5 and 5.5 dex, with a step of 0.5. All the models were for a solar
metallicity. The results of this procedure are displayed in
Figure~\ref{f:synthmodel}. From these results it is difficult to draw
a conclusion given that a lot of different models fit well with our results.
However, the results are compatible with a high $T_{eff}$ in the range between
3500 and 4000~K. Also, they seem to favour high surface gravity models.
These results are in contrast with what we obtained in Section~\ref{s:charcom}
where the value of $T_{eff}$ was slightly lower and moreover we were favouring
a low surface gravity result. 

\subsection{Characterization of the companion through SINFONI data}
\label{s:sinfonichar}

\begin{figure}
\centering
\includegraphics[width=\columnwidth]{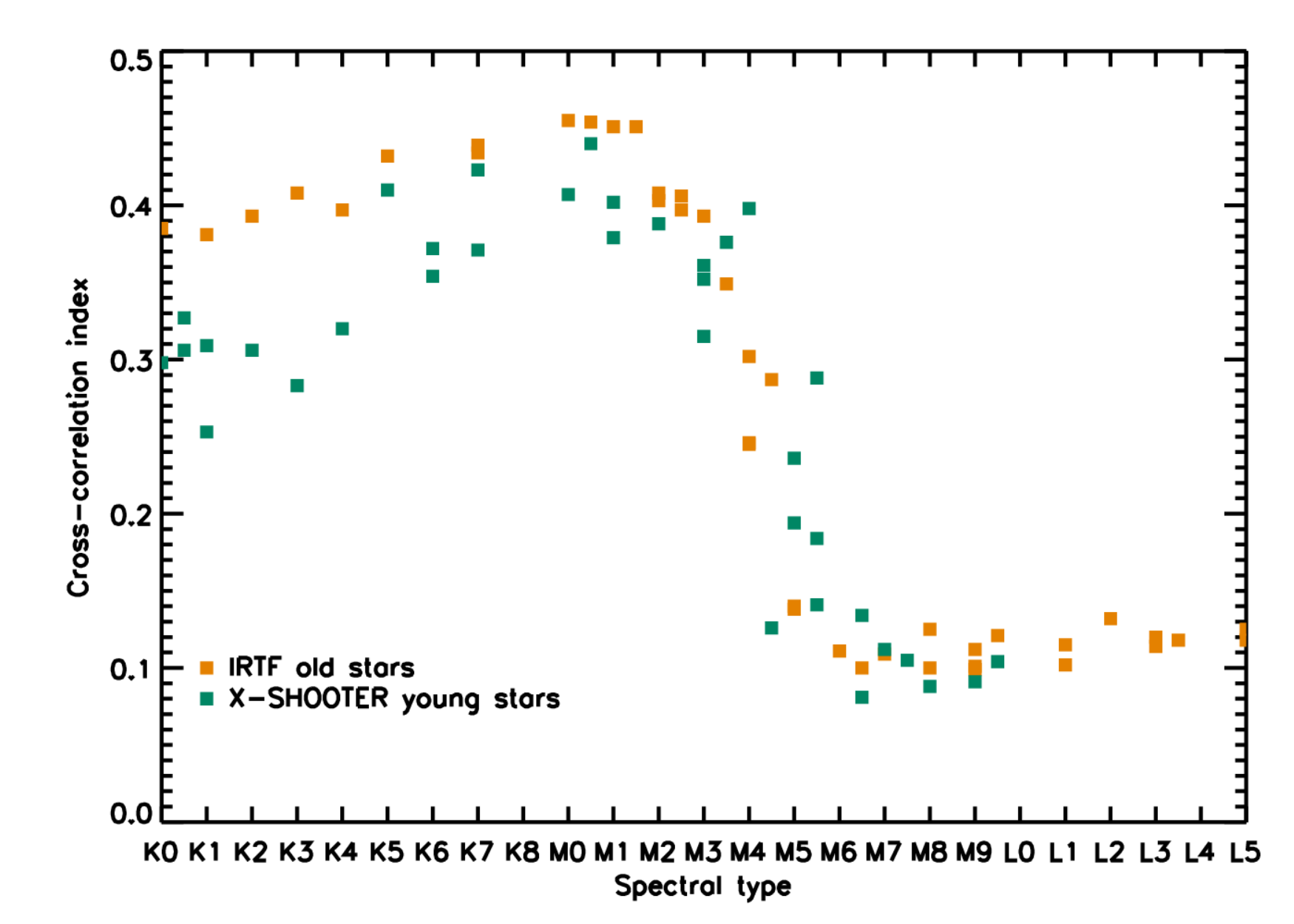}
\caption{Values of the cross-correlation index between the SINFONI spectrum
  of \rcra\,B and template spectra from the IRTF library (orange squares) and
  X-SHOOTER young stars (green squares) vs. the spectral type of the template
  spectra.}
\label{f:sinfosptype}
\end{figure}

\begin{figure}
\centering
\includegraphics[width=\columnwidth]{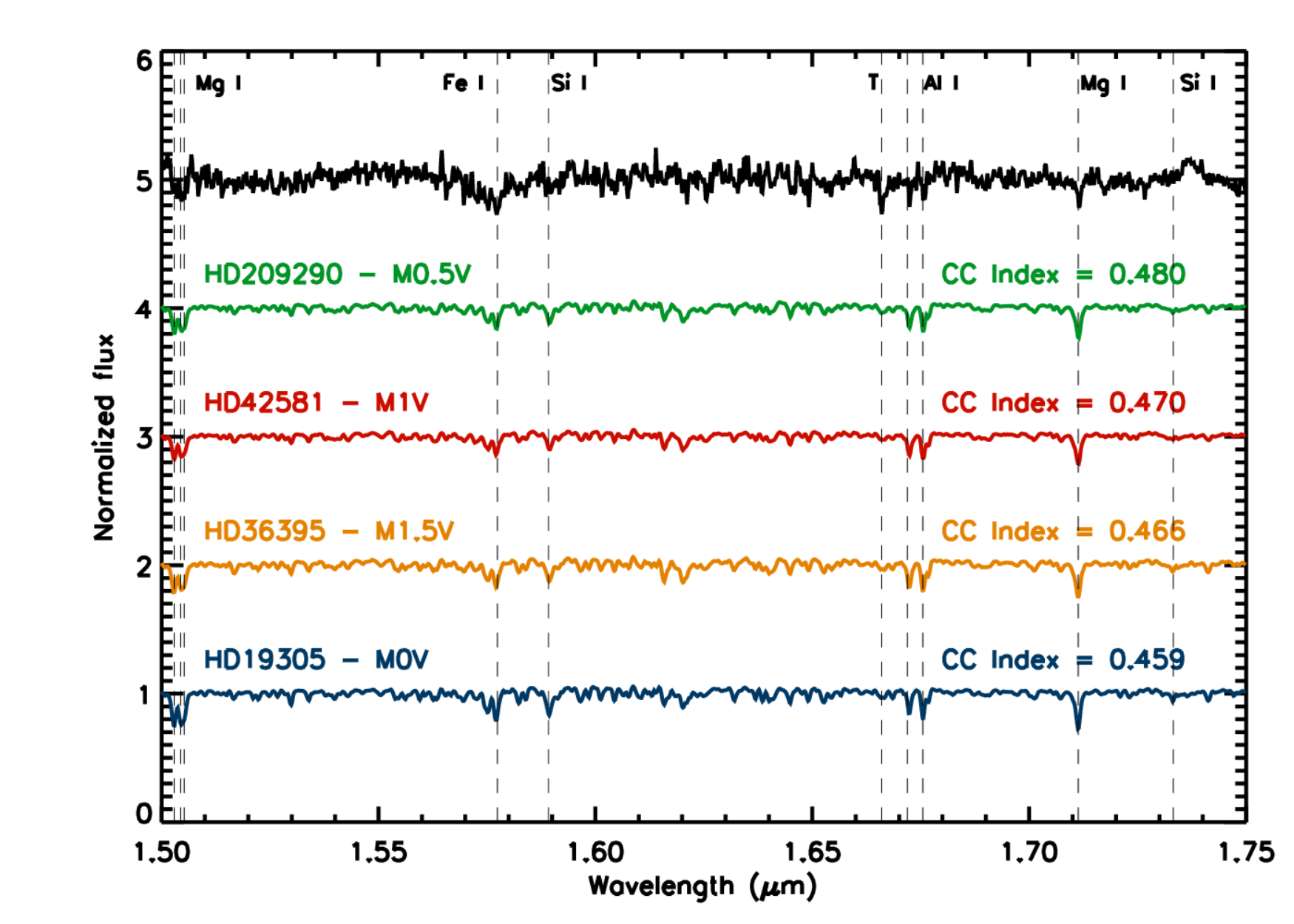}
\caption{Comparison of the \rcra\,B (black line) spectrum with those of some
  of the best fit spectra from the spectral libraries.}
\label{f:sinfocompa}
\end{figure}

The spectrum extracted from the SINFONI data was used to obtain a further
spectral classification of the object by comparing it with a library of
template spectra of K and M stars taken from the IRTF library
\citep{2005ApJ...623.1115C,2009ApJS..185..289R}. The comparison was done
through the spectral lines and, to avoid the final result to be influenced
by the slope of the spectra strongly dependent from the very
uncertain value of the extinction, we have divided each spectrum that we
used for a smoothed version of itself to eliminate any slope. The
smoothing was done with the standard SMOOTH IDL procedure using a
smoothing window of 100. The spectra were then compared using the
C\_CORRELATE IDL routine trying to maximize the cross-correlation index
between them. Moreover, to explore the possible shift of the spectra due
to the radial velocity of \rcra, this procedure was executed shifting the
template spectra of arbitrary wavelength values. In any case, we always
found that the highest cross-correlation indices were obtained for shifts
very near to zero concluding that the radial velocity of \rcra have to be
very close to 0.
The same procedure was then repeated using a library of
young stars spectra obtained by \citet{2013A&A...551A.107M} and
\citet{2017A&A...605A..86M} from X-SHOOTER data. In Figure~\ref{f:sinfosptype}
we display the values of the cross-correlation index as a function of the
spectral type that we obtain from the procedure described above showing in
orange the values obtained from the IRTF data and in green the values obtained
from the X-SHOOTER data. The evolution of the cross-correlation index is
very similar in both cases. The maximum values are
obtained for spectral types between M0 and M1.5 where the value of the
index is almost stable. On the low-mass side the values of the indices
diminish
with a steep slope while a shallower slope is found on the side of the K-stars.
In Figure~\ref{f:sinfocompa} we display the normalized spectrum of \rcra\,B
together with those of some of the best fit spectra from the IRTF library.
The fit of the spectral lines between the compared spectra is good with values
of the cross-correlation index between 0.46 and 0.48 for the best fit spectra.
In the image we have tagged some of the most prominent lines in the
spectra identifying lines from Mg~I, Fe~I, Si~I and Al~I. An evident line in
the \rcra\,B spectrum is also present at 1.67~\mic but it is not present in
the template spectra. It is probably a telluric line that has not been
completely deleted by the subtraction of the telluric spectrum standard
and for this reason we have tagged it with 'T' in Figure~\ref{f:sinfocompa}.
Using the spectral classification obtained from this
method, we obtain for \rcra\,B a $T_{eff}$ between 3650 and 3870~K that, from
the isochrones for pre-main sequence stars by \citet{2015A&A...577A..42B},
corresponds to a mass between 0.47 and 0.55 \MSun. \par
We also used the H lines in the spectral region observed with SINFONI to search
for any evidence of accretion. However, the upper limit to the equivalent width
(EW), obtained from the median of H lines in this region, is of 0.05~nm. As a
comparison we can consider the Orion low mass stars considered in
\citet{2012A&A...548A..56R} that have an EW of $\sim$0.1~nm. We can then
conclude that there is no evidence of accretion for \rcra\,B even if we have
to consider that accretion for this object is very probably variable while
we have obtained a spectrum in just one epoch.

\subsection{Orbital parameters}
\label{s:orbitpar}

\begin{figure*}[!htp]
\centering
\includegraphics[width=\textwidth]{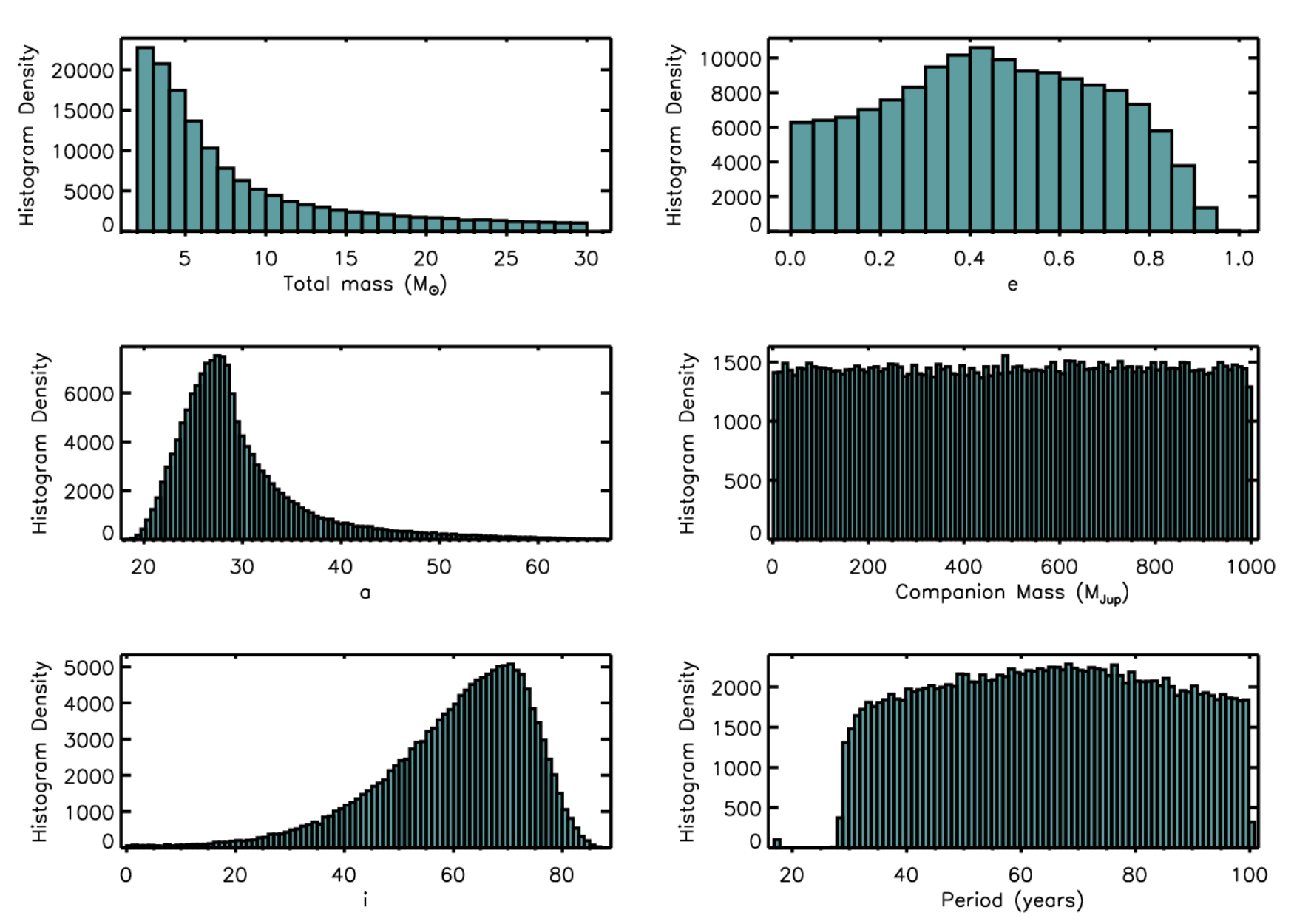}
\caption{Distribution histograms obtained from the Monte Carlo simulation
  described in Section~\ref{s:orbitpar} for the total mass of the system
  in solar masses {\it (upper left panel)}, the ellipticity {\it (upper right
    panel)}, the semi-major axis in au {\it (center left panel)}, the
  companion mass in \MJup {\it (center right panel)}, the inclination of the
  orbit {\it (bottom left panel)} and the period in years {\it bottom right
  panel)}.}
\label{f:histogram}
\end{figure*}

Using the astrometric data listed in Table~\ref{t:astro} we have performed a
Monte Carlo simulation to constrain the orbital parameters using the
Thiele-Innes formalism \citep{1960pdss.book.....B} as described in
\citet{2011A&A...533A..90D}, \citet{2013A&A...554A..21Z} and
\citet{2018MNRAS.480...35Z} and adopting the convention by
  \citet{2000eaa..bookE2855H}. The simulation generates by $5\times10^{7}$
random orbital elements and rejects all the central orbits that do not fit the
astrometric data. Given that the mass of the star is poorly constrained
through previous studies, we have also decided to include it between the
varying parameters of our simulation considering a very large mass range
of 2-30~\MSun. We also assumed a large variation of the mass of the companion
between 1 and 1000~\MJup. The simulation found 144249 orbits consistent
with the observational data. The results of this procedure for the main
parameters are shown in Figure~\ref{f:histogram}. The simulation clearly
prefers a low mass for the whole system, an eccentricity of $\sim$0.4, a
semi-major axis of 27-28~au and an inclination of $\sim70^{\circ}$. However, we
are not able to obtain a good value for the companion mass given that its
distribution is quite uniform and for the period even if in this case periods
of less than 30 years are clearly excluded and we have a very smoothed peak at
about 65-70 years. The availability of long term RV-data could help
in constraining the system mass-ratio. Unfortunately, we did not find any
data of this type for \rcra.

\subsection{Mass limits around \rcra}
\label{s:masslimit}

\begin{figure}
\centering
\includegraphics[width=\columnwidth]{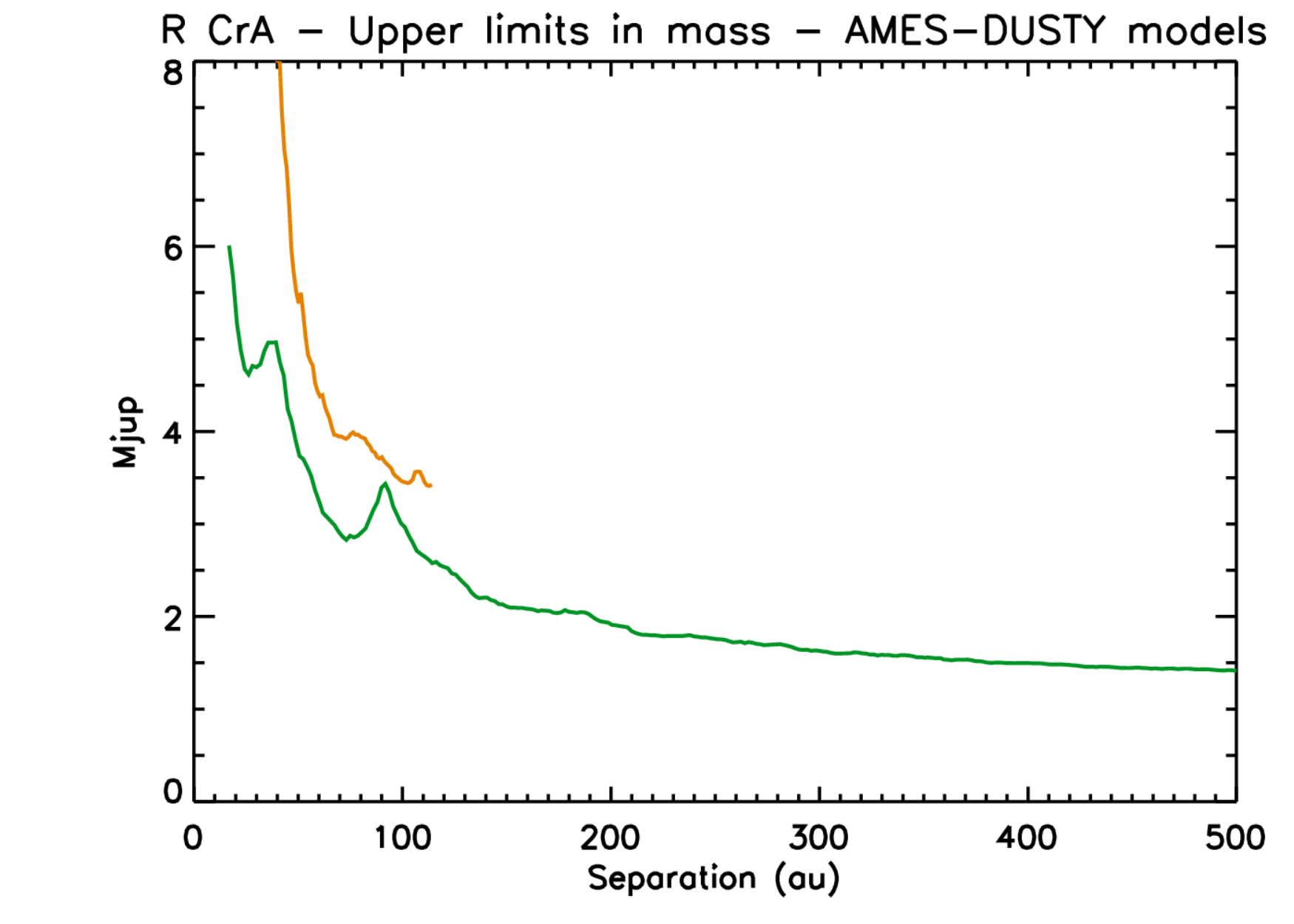}
\caption{Plot of the mass limits expressed in \MJup vs. the separation
  expressed in au for possible objects around \rcra. The IFS limits are given
  by the orange line while the IRDIS limits are given by the green line.}
\label{f:masslimit}
\end{figure}

With the aim to estimate the mass limits for other possible objects around
\rcra\ we have calculated the brightness contrast following the method devised
in \citet{mesa2015}. In this case, however, to overcome the problems arising
from the variability of \rcra, we have used as reference the star HIP\,63847
following the same procedure used to obtain the results in
Section~\ref{s:companion}. We then obtained the contrast in absolute
magnitude and we have corrected it for the extinction that we have calculated
for the companion in Table~\ref{t:photo}. While these values are not
necessarily valid for all the region around \rcra, we have used them to give
an estimate of the effective contrast. This result was then used in
association with the AMES-Dusty models to evaluate
the mass limits for other objects around the star. The results of this
procedure are displayed in Figure~\ref{f:masslimit}. Due to the much larger
extinction at shorter wavelengths, IFS is in this case less effective in
finding low mass companions, hence we can use IRDIS to set these limits.
At separations from the star less than 100~au, we are able to obtain mass limits
variable, according to the separation, between 6 and 2.5~\MJup while at larger
separations we can reach an almost stable mass limit of the order of 1.5~\MJup. 

\section{Conclusions}
\label{s:conclusion}

In this paper we report the results from the SPHERE multi-epoch observations
of the \rcra system. The central star is a very young ($\sim$1~Myr) HAeBe
object with a strong reddening due to the presence of circumstellar material.
The star is strongly variable on period of $\sim$66~days and, moreover, it has
strong differences in magnitude between different spectral bands. These
facts make difficult a proper photometric characterization of companion
objects. \par
A companion was found around the star and, thanks to the astrometric
measurements taken during the four observing epochs, we were able to confirm
that it is gravitationally bound to the star. Also, we were able to extract a
spectrum of the companion. After correction for extinction in
the direction of the companion and through the comparison with the BT-Settl
evolutionary models we were able to infer for the companion a mass of
0.29$\pm$0.08~\MSun.
Hence, the companion is an early M spectral type star deeply embedded in its
dust envelope. This case is then very similar to that of the two
companions of T\,Tau, T\,Tau\,S\,a and T\,Tau\,S\,b
\citep{1982ApJ...255L.103D,2016A&A...593A..50K}. An alternative explanation
of the strong extinction in the direction of the companion is that it could
be due to the presence of an external disk seen at very large inclination.
Despite the fact that we have found some hints of the presence of this external
disk, we are not able at the moment to disentangle between the two explanations
of the large extinction in the direction of the companion. The spectral
classification of the companion is further confirmed
comparing template spectra to the companion spectrum obtained using SINFONI.
The results obtained with the latter instrument provide a slightly higher value
both for the mass (0.47-0.55~\MSun) and for $T_{eff}$ (3650-3870~K) than what
obtained from the SPHERE data. This discrepancy could be due to the fact that
the extinction value determined from SPHERE data is slightly underestimated
obtaining in this latter case a value of $A_J$=5.00 mag. Moreover, given that
the classification through the
SINFONI data relies on spectral lines that are not affected by the extinction,
we believe that the latter are a more reliable estimate of the companion
mass. The higher temperature derived from the spectral lines through SINFONI
data might be due to the presence of a hotter region on the stellar surface
related to accretion. In addition, we notice that no
emission line was detected in our SINFONI spectrum, although several lines in
the Brackett series are included in the observed spectral range with an upper
limit to the EW of 0.05~nm. It is then worth to notice that, using
for the star the distance of 94.9~pc derived from the Gaia parallax, we would
obtain from the comparison with the BT-Settl evolutionary models a mass of the
order of $0.14\pm0.03$~\MSun corresponding to a spectral type of M5-M6. This
would be even more in contrast with the mass determination obtained from
the SINFONI spectrum of the companion that is not depending from the
distance of the star. This gives even more strength to our estimate of the
distance of the system made in Section~\ref{s:star}.\par
The position of the stellar companion probably in a gap in the disk
around the primary star of the system could make this object a candidate
for formation through disk instability as proposed e.g. by
\citet{2016MNRAS.463..957F} for massive young protostars with a large
disc-star mass ratio. A similar object has been recently discovered
through ALMA observations around the young stellar object G11.92-0.61 MM1
\citep{2018arXiv181105267I}. In any case, to confirm this possibility we will
need more precise informations both about the physical characteristics (e.g.
its mass that is still poorly defined) of the primary star of the system and
on the effective location of the companion into the gap of the disk for which
we have at the moment just non-conclusive hints. \par
Aside to the presence of the companion, the environment around \rcra\ is
further enriched by the presence of a number of extended structures. The
most notable of them is surely the bright jet-like structure North-East of the
star. Our images highlight moreover that this jet-like structure is itself
formed by at least two substructures. A less bright counter-jet-like structure
on the opposite side of the star is also visible together with some arc-like
structures that could be part of an external disk. Instead, in our images it
is not possible to image the inner disk that the star is known to host. We will
discuss the nature and the origin of the jet-like structure and of all the
other extended structures in this system in a following dedicated paper.


\begin{acknowledgements}
The authors thanks the anonymous referee for the constructive comments that
helped to strongly improve the quality of the present work. \par
This work has made use of the SPHERE Data Center, jointly operated by
OSUG/IPAG (Grenoble), PYTHEAS/LAM/CeSAM (Marseille), OCA/Lagrange (Nice) and
Observatoire de Paris/LESIA (Paris). \par
This work has made use of data from the European Space Agency (ESA) mission
{\it Gaia} (\url{https://www.cosmos.esa.int/gaia}), processed by
the {\it Gaia} Data Processing and Analysis Consortium (DPAC,
\url{https://www.cosmos.esa.int/web/gaia/dpac/consortium}). Funding for
the DPAC has been provided by national institutions, in particular the
institutions participating in the {\it Gaia} Multilateral Agreement. \par
This research has made use of the SIMBAD database, operated at CDS,
Strasbourg, France. \par
D.M. acknowledges support from the ESO-Government of Chile Joint Comittee
program 'Direct imaging and characterization of exoplanets'. 
D.M., A.Z., V.D.O., R.G., R.U.C., S.D., C.L. acknowledge support from
the ``Progetti Premiali'' funding scheme of the Italian Ministry of Education,
University, and Research. A.Z. acknowledges support from the CONICYT + PAI/
Convocatoria nacional subvenci\'on a la instalaci\'on en la academia,
convocatoria 2017 + Folio PAI77170087. M.G.U.G. and G.L. acknowledge support
from the project PRIN-INAF 2016 The Cradle of Life - GENESIS-SKA (General
Conditions in Early Planetary Systems for the rise of life with SKA). G.P. and
M.L. acknowledge financial support from the ANR of France under contract number
ANR-16-CE31-0013 (Planet-Forming-Disks). D.F. acknowledge financial support
provided by the Italian Ministry of Education, Universities and Research,
project SIR (RBSI14ZRHR). \par
This publication makes use of VOSA, developed under the Spanish Virtual
Observatory project supported from the Spanish MINECO through grant
AyA2017-84089. This work has been supported by a grant from the Agence
Nationale de la Recherche (grant ANR-14-CE33-0018). \par
SPHERE is an instrument designed and built by a consortium consisting
of IPAG (Grenoble, France), MPIA (Heidelberg, Germany), LAM (Marseille,
France), LESIA (Paris, France), Laboratoire Lagrange (Nice, France),
INAF-Osservatorio di Padova (Italy), Observatoire de Gen\`eve (Switzerland),
ETH Zurich (Switzerland), NOVA (Netherlands), ONERA (France) and ASTRON
(Netherlands), in collaboration with ESO. SPHERE was funded by ESO, with
additional contributions from CNRS (France), MPIA (Germany), INAF (Italy),
FINES (Switzerland) and NOVA (Netherlands). SPHERE also received funding
from the European Commission Sixth and Seventh Framework Programmes as
part of the Optical Infrared Coordination Network for Astronomy (OPTICON)
under grant number RII3-Ct-2004-001566 for FP6 (2004-2008), grant number
226604 for FP7 (2009-2012) and grant number 312430 for FP7 (2013-2016).
\end{acknowledgements}

\bibliographystyle{aa}
\bibliography{RCrA_paper}

\begin{thebibliography}{82}
\expandafter\ifx\csname natexlab\endcsname\relax\def\natexlab#1{#1}\fi

\bibitem[{{Allard}(2014)}]{2014IAUS..299..271A}
{Allard}, F. 2014, in IAU Symposium, Vol. 299, Exploring the Formation and
  Evolution of Planetary Systems, ed. M.~{Booth}, B.~C. {Matthews}, \& J.~R.
  {Graham}, 271--272

\bibitem[{{Allard} {et~al.}(1997){Allard}, {Hauschildt}, {Alexander}, \&
  {Starrfield}}]{1997ARA&A..35..137A}
{Allard}, F., {Hauschildt}, P.~H., {Alexander}, D.~R., \& {Starrfield}, S.
  1997, \araa, 35, 137

\bibitem[{{Allard} {et~al.}(2012){Allard}, {Homeier}, \&
  {Freytag}}]{2012RSPTA.370.2765A}
{Allard}, F., {Homeier}, D., \& {Freytag}, B. 2012, Philosophical Transactions
  of the Royal Society of London Series A, 370, 2765

\bibitem[{{Allers} \& {Liu}(2013)}]{2013ApJ...772...79A}
{Allers}, K.~N. \& {Liu}, M.~C. 2013, \apj, 772, 79

\bibitem[{{Baraffe} {et~al.}(2015){Baraffe}, {Homeier}, {Allard}, \&
  {Chabrier}}]{2015A&A...577A..42B}
{Baraffe}, I., {Homeier}, D., {Allard}, F., \& {Chabrier}, G. 2015, \aap, 577,
  A42

\bibitem[{{Bayo} {et~al.}(2008){Bayo}, {Rodrigo}, {Barrado Y Navascu{\'e}s},
  {Solano}, {Guti{\'e}rrez}, {Morales-Calder{\'o}n}, \&
  {Allard}}]{2008A&A...492..277B}
{Bayo}, A., {Rodrigo}, C., {Barrado Y Navascu{\'e}s}, D., {et~al.} 2008, \aap,
  492, 277

\bibitem[{{Beuzit} {et~al.}(2008){Beuzit}, {Feldt}, {Dohlen}, {Mouillet},
  {Puget}, {Wildi}, {Abe}, {Antichi}, {Baruffolo}, {Baudoz}, {Boccaletti},
  {Carbillet}, {Charton}, {Claudi}, {Downing}, {Fabron}, {Feautrier},
  {Fedrigo}, {Fusco}, {Gach}, {Gratton}, {Henning}, {Hubin}, {Joos}, {Kasper},
  {Langlois}, {Lenzen}, {Moutou}, {Pavlov}, {Petit}, {Pragt}, {Rabou}, {Rigal},
  {Roelfsema}, {Rousset}, {Saisse}, {Schmid}, {Stadler}, {Thalmann}, {Turatto},
  {Udry}, {Vakili}, \& {Waters}}]{2008SPIE.7014E..18B}
{Beuzit}, J.-L., {Feldt}, M., {Dohlen}, K., {et~al.} 2008, in \procspie, Vol.
  7014, Ground-based and Airborne Instrumentation for Astronomy II, 701418

\bibitem[{{Bibo} {et~al.}(1992){Bibo}, {The}, \&
  {Dawanas}}]{1992A&A...260..293B}
{Bibo}, E.~A., {The}, P.~S., \& {Dawanas}, D.~N. 1992, \aap, 260, 293

\bibitem[{{Binnendijk}(1960)}]{1960pdss.book.....B}
{Binnendijk}, L. 1960, {Properties of double stars; a survey of parallaxes and
  orbits.}

\bibitem[{{Bonnet} {et~al.}(2004){Bonnet}, {Conzelmann}, {Delabre},
  {Donaldson}, {Fedrigo}, {Hubin}, {Kissler-Patig}, {Lizon}, {Paufique},
  {Rossi}, {Stroebele}, \& {Tordo}}]{2004SPIE.5490..130B}
{Bonnet}, H., {Conzelmann}, R., {Delabre}, B., {et~al.} 2004, in \procspie,
  Vol. 5490, Advancements in Adaptive Optics, ed. D.~{Bonaccini Calia}, B.~L.
  {Ellerbroek}, \& R.~{Ragazzoni}, 130--138

\bibitem[{{Bowler}(2016)}]{2016PASP..128j2001B}
{Bowler}, B.~P. 2016, \pasp, 128, 102001

\bibitem[{{Burgasser}(2014)}]{2014ASInC..11....7B}
{Burgasser}, A.~J. 2014, in Astronomical Society of India Conference Series,
  Vol.~11, Astronomical Society of India Conference Series

\bibitem[{{Cardelli} {et~al.}(1989){Cardelli}, {Clayton}, \&
  {Mathis}}]{1989ApJ...345..245C}
{Cardelli}, J.~A., {Clayton}, G.~C., \& {Mathis}, J.~S. 1989, \apj, 345, 245

\bibitem[{{Chauvin} {et~al.}(2017){Chauvin}, {Desidera}, {Lagrange}, {Vigan},
  {Feldt}, {Gratton}, {Langlois}, {Cheetham}, {Bonnefoy}, \&
  {Meyer}}]{2017sf2a.conf..331C}
{Chauvin}, G., {Desidera}, S., {Lagrange}, A.-M., {et~al.} 2017, in SF2A-2017:
  Proceedings of the Annual meeting of the French Society of Astronomy and
  Astrophysics, ed. C.~{Reyl{\'e}}, P.~{Di Matteo}, F.~{Herpin}, E.~{Lagadec},
  A.~{Lan{\c c}on}, Z.~{Meliani}, \& F.~{Royer}, 331--335

\bibitem[{{Chen} {et~al.}(2012){Chen}, {Pecaut}, {Mamajek}, {Su}, \&
  {Bitner}}]{2012ApJ...756..133C}
{Chen}, C.~H., {Pecaut}, M., {Mamajek}, E.~E., {Su}, K.~Y.~L., \& {Bitner}, M.
  2012, \apj, 756, 133

\bibitem[{{Chen} {et~al.}(1997){Chen}, {Grenfell}, {Myers}, {P.~C.}, \&
  {Hughes}}]{1997ApJ...478..295C}
{Chen}, H., {Grenfell}, T.~G., {Myers}, {P.~C.}, \& {Hughes}, J.~D. 1997, \apj,
  478, 295

\bibitem[{{Clark} {et~al.}(2000){Clark}, {McCall}, {Chrysostomou}, {Gledhill},
  {Yates}, \& {Hough}}]{2000MNRAS.319..337C}
{Clark}, S., {McCall}, A., {Chrysostomou}, A., {et~al.} 2000, \mnras, 319, 337

\bibitem[{{Claudi} {et~al.}(2008){Claudi}, {Turatto}, {Gratton}, {Antichi},
  {Bonavita}, {Bruno}, {Cascone}, {De Caprio}, {Desidera}, {Giro}, {Mesa},
  {Scuderi}, {Dohlen}, {Beuzit}, \& {Puget}}]{Cl08}
{Claudi}, R.~U., {Turatto}, M., {Gratton}, R.~G., {et~al.} 2008, in Society of
  Photo-Optical Instrumentation Engineers (SPIE) Conference Series, Vol. 7014,
  Society of Photo-Optical Instrumentation Engineers (SPIE) Conference Series

\bibitem[{{Cushing} {et~al.}(2005){Cushing}, {Rayner}, \&
  {Vacca}}]{2005ApJ...623.1115C}
{Cushing}, M.~C., {Rayner}, J.~T., \& {Vacca}, W.~D. 2005, \apj, 623, 1115

\bibitem[{{Cutri} {et~al.}(2003){Cutri}, {Skrutskie}, {van Dyk}, {Beichman},
  {Carpenter}, {Chester}, {Cambresy}, {Evans}, {Fowler}, {Gizis}, {Howard},
  {Huchra}, {Jarrett}, {Kopan}, {Kirkpatrick}, {Light}, {Marsh}, {McCallon},
  {Schneider}, {Stiening}, {Sykes}, {Weinberg}, {Wheaton}, {Wheelock}, \&
  {Zacarias}}]{2003yCat.2246....0C}
{Cutri}, R.~M., {Skrutskie}, M.~F., {van Dyk}, S., {et~al.} 2003, VizieR Online
  Data Catalog, 2246

\bibitem[{{de Zeeuw} {et~al.}(1999){de Zeeuw}, {Hoogerwerf}, {de Bruijne},
  {Brown}, \& {Blaauw}}]{1999AJ....117..354D}
{de Zeeuw}, P.~T., {Hoogerwerf}, R., {de Bruijne}, J.~H.~J., {Brown}, A.~G.~A.,
  \& {Blaauw}, A. 1999, \aj, 117, 354

\bibitem[{{Delorme} {et~al.}(2017){Delorme}, {Meunier}, {Albert}, {Lagadec},
  {Le Coroller}, {Galicher}, {Mouillet}, {Boccaletti}, {Mesa}, {Meunier},
  {Beuzit}, {Lagrange}, {Chauvin}, {Sapone}, {Langlois}, {Maire},
  {Montarg{\`e}s}, {Gratton}, {Vigan}, \& {Surace}}]{2017sf2a.conf..347D}
{Delorme}, P., {Meunier}, N., {Albert}, D., {et~al.} 2017, in SF2A-2017:
  Proceedings of the Annual meeting of the French Society of Astronomy and
  Astrophysics, ed. C.~{Reyl{\'e}}, P.~{Di Matteo}, F.~{Herpin}, E.~{Lagadec},
  A.~{Lan{\c c}on}, Z.~{Meliani}, \& F.~{Royer}, 347--361

\bibitem[{{Desidera} {et~al.}(2011){Desidera}, {Carolo}, {Gratton}, {Martinez
  Fiorenzano}, {Endl}, {Mesa}, {Barbieri}, {Bonavita}, {Cecconi}, {Claudi},
  {Cosentino}, {Marzari}, \& {Scuderi}}]{2011A&A...533A..90D}
{Desidera}, S., {Carolo}, E., {Gratton}, R., {et~al.} 2011, \aap, 533, A90

\bibitem[{{Dohlen} {et~al.}(2008){Dohlen}, {Langlois}, {Saisse}, {Hill},
  {Origne}, {Jacquet}, {Fabron}, {Blanc}, {Llored}, {Carle}, {Moutou}, {Vigan},
  {Boccaletti}, {Carbillet}, {Mouillet}, \& {Beuzit}}]{Do08}
{Dohlen}, K., {Langlois}, M., {Saisse}, M., {et~al.} 2008, in Society of
  Photo-Optical Instrumentation Engineers (SPIE) Conference Series, Vol. 7014,
  Society of Photo-Optical Instrumentation Engineers (SPIE) Conference Series

\bibitem[{{Ducati}(2002)}]{2002yCat.2237....0D}
{Ducati}, J.~R. 2002, VizieR Online Data Catalog, 2237

\bibitem[{{Dyck} {et~al.}(1982){Dyck}, {Simon}, \&
  {Zuckerman}}]{1982ApJ...255L.103D}
{Dyck}, H.~M., {Simon}, T., \& {Zuckerman}, B. 1982, \apjl, 255, L103

\bibitem[{{Dzib} {et~al.}(2018){Dzib}, {Loinard}, {Ortiz-Le{\'o}n},
  {Rodr{\'{\i}}guez}, \& {Galli}}]{2018ApJ...867..151D}
{Dzib}, S.~A., {Loinard}, L., {Ortiz-Le{\'o}n}, G.~N., {Rodr{\'{\i}}guez},
  L.~F., \& {Galli}, P.~A.~B. 2018, \apj, 867, 151

\bibitem[{{Eisenhauer} {et~al.}(2003){Eisenhauer}, {Abuter}, {Bickert},
  {Biancat-Marchet}, {Bonnet}, {Brynnel}, {Conzelmann}, {Delabre}, {Donaldson},
  {Farinato}, {Fedrigo}, {Genzel}, {Hubin}, {Iserlohe}, {Kasper},
  {Kissler-Patig}, {Monnet}, {Roehrle}, {Schreiber}, {Stroebele}, {Tecza},
  {Thatte}, \& {Weisz}}]{2003SPIE.4841.1548E}
{Eisenhauer}, F., {Abuter}, R., {Bickert}, K., {et~al.} 2003, in \procspie,
  Vol. 4841, Instrument Design and Performance for Optical/Infrared
  Ground-based Telescopes, ed. M.~{Iye} \& A.~F.~M. {Moorwood}, 1548--1561

\bibitem[{{Forbrich} {et~al.}(2006){Forbrich}, {Preibisch}, \&
  {Menten}}]{2006A&A...446..155F}
{Forbrich}, J., {Preibisch}, T., \& {Menten}, K.~M. 2006, \aap, 446, 155

\bibitem[{{Forgan} {et~al.}(2016){Forgan}, {Ilee}, {Cyganowski}, {Brogan}, \&
  {Hunter}}]{2016MNRAS.463..957F}
{Forgan}, D.~H., {Ilee}, J.~D., {Cyganowski}, C.~J., {Brogan}, C.~L., \&
  {Hunter}, T.~R. 2016, \mnras, 463, 957

\bibitem[{{Gaia Collaboration}(2018)}]{2018yCat.1345....0G}
{Gaia Collaboration}. 2018, VizieR Online Data Catalog, 1345

\bibitem[{{Galicher} {et~al.}(2018){Galicher}, {Boccaletti}, {Mesa}, {Delorme},
  {Gratton}, {Langlois}, {Lagrange}, {Maire}, {Le Coroller}, {Chauvin},
  {Biller}, {Cantalloube}, {Janson}, {Lagadec}, {Meunier}, {Vigan},
  {Hagelberg}, {Bonnefoy}, {Zurlo}, {Rocha}, {Maurel}, {Jaquet}, {Buey}, \&
  {Weber}}]{2018A&A...615A..92G}
{Galicher}, R., {Boccaletti}, A., {Mesa}, D., {et~al.} 2018, \aap, 615, A92

\bibitem[{{Garcia Lopez} {et~al.}(2006){Garcia Lopez}, {Natta}, {Testi}, \&
  {Habart}}]{2006A&A...459..837G}
{Garcia Lopez}, R., {Natta}, A., {Testi}, L., \& {Habart}, E. 2006, \aap, 459,
  837

\bibitem[{{Graham}(1992)}]{1992lmsf.book..185G}
{Graham}, J.~A. 1992, {Star Formation in the Corona Australis Region}, ed.
  B.~{Reipurth}, 185

\bibitem[{{Graham}(1993)}]{1993PASP..105..561G}
{Graham}, J.~A. 1993, \pasp, 105, 561

\bibitem[{{Gray} {et~al.}(2006){Gray}, {Corbally}, {Garrison}, {McFadden},
  {Bubar}, {McGahee}, {O'Donoghue}, \& {Knox}}]{2006AJ....132..161G}
{Gray}, R.~O., {Corbally}, C.~J., {Garrison}, R.~F., {et~al.} 2006, \aj, 132,
  161

\bibitem[{{Heintz}(2000)}]{2000eaa..bookE2855H}
{Heintz}, W. 2000, {Visual Binary Stars}, ed. P.~{Murdin}, 2855

\bibitem[{{Herbig}(1960)}]{1960ApJS....4..337H}
{Herbig}, G.~H. 1960, \apjs, 4, 337

\bibitem[{{Herbst} \& {Shevchenko}(1999)}]{1999AJ....118.1043H}
{Herbst}, W. \& {Shevchenko}, V.~S. 1999, \aj, 118, 1043

\bibitem[{{Hillenbrand} {et~al.}(1992){Hillenbrand}, {Strom}, {Vrba}, \&
  {Keene}}]{1992ApJ...397..613H}
{Hillenbrand}, L.~A., {Strom}, S.~E., {Vrba}, F.~J., \& {Keene}, J. 1992, \apj,
  397, 613

\bibitem[{{Hoeijmakers} {et~al.}(2018){Hoeijmakers}, {Schwarz}, {Snellen}, {de
  Kok}, {Bonnefoy}, {Chauvin}, {Lagrange}, \& {Girard}}]{2018A&A...617A.144H}
{Hoeijmakers}, H.~J., {Schwarz}, H., {Snellen}, I.~A.~G., {et~al.} 2018, \aap,
  617, A144

\bibitem[{{Ilee} {et~al.}(2018){Ilee}, {Cyganowski}, {Brogan}, {Hunter},
  {Forgan}, {Haworth}, {Clarke}, \& {Harries}}]{2018arXiv181105267I}
{Ilee}, J.~D., {Cyganowski}, C.~J., {Brogan}, C.~L., {et~al.} 2018, ArXiv
  e-prints

\bibitem[{{Johnson} {et~al.}(2010){Johnson}, {Aller}, {Howard}, \&
  {Crepp}}]{2010PASP..122..905J}
{Johnson}, J.~A., {Aller}, K.~M., {Howard}, A.~W., \& {Crepp}, J.~R. 2010,
  \pasp, 122, 905

\bibitem[{{Kasper} {et~al.}(2016){Kasper}, {Santhakumari}, {Herbst}, \&
  {K{\"o}hler}}]{2016A&A...593A..50K}
{Kasper}, M., {Santhakumari}, K.~K.~R., {Herbst}, T.~M., \& {K{\"o}hler}, R.
  2016, \aap, 593, A50

\bibitem[{{Koen} {et~al.}(2010){Koen}, {Kilkenny}, {van Wyk}, \&
  {Marang}}]{2010MNRAS.403.1949K}
{Koen}, C., {Kilkenny}, D., {van Wyk}, F., \& {Marang}, F. 2010, \mnras, 403,
  1949

\bibitem[{{Kraus} {et~al.}(2009){Kraus}, {Hofmann}, {Malbet}, {Meilland},
  {Natta}, {Schertl}, {Stee}, \& {Weigelt}}]{2009A&A...508..787K}
{Kraus}, S., {Hofmann}, K.-H., {Malbet}, F., {et~al.} 2009, \aap, 508, 787

\bibitem[{{Langlois} {et~al.}(2013){Langlois}, {Vigan}, {Moutou}, {Sauvage},
  {Dohlen}, {Costille}, {Mouillet}, \& {Le Mignant}}]{2013aoel.confE..63L}
{Langlois}, M., {Vigan}, A., {Moutou}, C., {et~al.} 2013, in Proceedings of the
  Third AO4ELT Conference, ed. S.~{Esposito} \& L.~{Fini}, 63

\bibitem[{{Lazareff} {et~al.}(2017){Lazareff}, {Berger}, {Kluska}, {Le
  Bouquin}, {Benisty}, {Malbet}, {Koen}, {Pinte}, {Thi}, {Absil}, {Baron},
  {Delboulb{\'e}}, {Duvert}, {Isella}, {Jocou}, {Juhasz}, {Kraus}, {Lachaume},
  {M{\'e}nard}, {Millan-Gabet}, {Monnier}, {Moulin}, {Perraut}, {Rochat},
  {Soulez}, {Tallon}, {Thi{\'e}baut}, {Traub}, \& {Zins}}]{2017A&A...599A..85L}
{Lazareff}, B., {Berger}, J.-P., {Kluska}, J., {et~al.} 2017, \aap, 599, A85

\bibitem[{{Ligi} {et~al.}(2018){Ligi}, {Demangeon}, {Barros}, {Mesa},
  {Bonavita}, {Vigan}, {Bonnefoy}, {Gratton}, \&
  {Deleuil}}]{2018AJ....156..182L}
{Ligi}, R., {Demangeon}, O., {Barros}, S., {et~al.} 2018, \aj, 156, 182

\bibitem[{{Malfait} {et~al.}(1998){Malfait}, {Bogaert}, \&
  {Waelkens}}]{1998A&A...331..211M}
{Malfait}, K., {Bogaert}, E., \& {Waelkens}, C. 1998, \aap, 331, 211

\bibitem[{{Manara} {et~al.}(2017){Manara}, {Frasca}, {Alcal{\'a}}, {Natta},
  {Stelzer}, \& {Testi}}]{2017A&A...605A..86M}
{Manara}, C.~F., {Frasca}, A., {Alcal{\'a}}, J.~M., {et~al.} 2017, \aap, 605,
  A86

\bibitem[{{Manara} {et~al.}(2013){Manara}, {Testi}, {Rigliaco}, {Alcal{\'a}},
  {Natta}, {Stelzer}, {Biazzo}, {Covino}, {Covino}, {Cupani}, {D'Elia}, \&
  {Randich}}]{2013A&A...551A.107M}
{Manara}, C.~F., {Testi}, L., {Rigliaco}, E., {et~al.} 2013, \aap, 551, A107

\bibitem[{{Marois} {et~al.}(2014){Marois}, {Correia}, {Galicher}, {Ingraham},
  {Macintosh}, {Currie}, \& {De Rosa}}]{2014SPIE.9148E..0UM}
{Marois}, C., {Correia}, C., {Galicher}, R., {et~al.} 2014, in \procspie, Vol.
  9148, Adaptive Optics Systems IV, 91480U

\bibitem[{{Marois} {et~al.}(2006{\natexlab{a}}){Marois}, {Lafreni{\`e}re},
  {Doyon}, {Macintosh}, \& {Nadeau}}]{2006ApJ...641..556M}
{Marois}, C., {Lafreni{\`e}re}, D., {Doyon}, R., {Macintosh}, B., \& {Nadeau},
  D. 2006{\natexlab{a}}, \apj, 641, 556

\bibitem[{{Marois} {et~al.}(2006{\natexlab{b}}){Marois}, {Lafreni{\`e}re},
  {Macintosh}, \& {Doyon}}]{2006ApJ...647..612M}
{Marois}, C., {Lafreni{\`e}re}, D., {Macintosh}, B., \& {Doyon}, R.
  2006{\natexlab{b}}, \apj, 647, 612

\bibitem[{{Marshall} {et~al.}(2014){Marshall}, {Moro-Mart{\'{\i}}n}, {Eiroa},
  {Kennedy}, {Mora}, {Sibthorpe}, {Lestrade}, {Maldonado}, {Sanz-Forcada},
  {Wyatt}, {Matthews}, {Horner}, {Montesinos}, {Bryden}, {del Burgo},
  {Greaves}, {Ivison}, {Meeus}, {Olofsson}, {Pilbratt}, \&
  {White}}]{2014A&A...565A..15M}
{Marshall}, J.~P., {Moro-Mart{\'{\i}}n}, A., {Eiroa}, C., {et~al.} 2014, \aap,
  565, A15

\bibitem[{{Mesa} {et~al.}(2018){Mesa}, {Baudino}, {Charnay}, {D'Orazi},
  {Desidera}, {Boccaletti}, {Gratton}, {Bonnefoy}, {Delorme}, {Langlois},
  {Vigan}, {Zurlo}, {Maire}, {Janson}, {Antichi}, {Baruffolo}, {Bruno},
  {Cascone}, {Chauvin}, {Claudi}, {De Caprio}, {Fantinel}, {Farisato}, {Feldt},
  {Giro}, {Hagelberg}, {Incorvaia}, {Lagadec}, {Lagrange}, {Lazzoni}, {Lessio},
  {Salasnich}, {Scuderi}, {Sissa}, \& {Turatto}}]{2018A&A...612A..92M}
{Mesa}, D., {Baudino}, J.-L., {Charnay}, B., {et~al.} 2018, \aap, 612, A92

\bibitem[{{Mesa} {et~al.}(2015){Mesa}, {Gratton}, {Zurlo}, {Vigan}, {Claudi},
  {Alberi}, {Antichi}, {Baruffolo}, {Beuzit}, {Boccaletti}, {Bonnefoy},
  {Costille}, {Desidera}, {Dohlen}, {Fantinel}, {Feldt}, {Fusco}, {Giro},
  {Henning}, {Kasper}, {Langlois}, {Maire}, {Martinez}, {Moeller-Nilsson},
  {Mouillet}, {Moutou}, {Pavlov}, {Puget}, {Salasnich}, {Sauvage}, {Sissa},
  {Turatto}, {Udry}, {Vakili}, {Waters}, \& {Wildi}}]{mesa2015}
{Mesa}, D., {Gratton}, R., {Zurlo}, A., {et~al.} 2015, \aap, 576, A121

\bibitem[{{Mesa} {et~al.}(2016){Mesa}, {Vigan}, {D'Orazi}, {Ginski},
  {Desidera}, {Bonnefoy}, {Gratton}, {Langlois}, {Marzari}, {Messina},
  {Antichi}, {Biller}, {Bonavita}, {Cascone}, {Chauvin}, {Claudi}, {Curtis},
  {Fantinel}, {Feldt}, {Garufi}, {Galicher}, {Henning}, {Incorvaia},
  {Lagrange}, {Millward}, {Perrot}, {Salasnich}, {Scuderi}, {Sissa}, {Wahhaj},
  \& {Zurlo}}]{2016A&A...593A.119M}
{Mesa}, D., {Vigan}, A., {D'Orazi}, V., {et~al.} 2016, \aap, 593, A119

\bibitem[{{Meshkat} {et~al.}(2015){Meshkat}, {Bonnefoy}, {Mamajek}, {Quanz},
  {Chauvin}, {Kenworthy}, {Rameau}, {Meyer}, {Lagrange}, {Lannier}, \&
  {Delorme}}]{2015MNRAS.453.2378M}
{Meshkat}, T., {Bonnefoy}, M., {Mamajek}, E.~E., {et~al.} 2015, \mnras, 453,
  2378

\bibitem[{{Meyer} \& {Wilking}(2009)}]{2009PASP..121..350M}
{Meyer}, M.~R. \& {Wilking}, B.~A. 2009, \pasp, 121, 350

\bibitem[{{Natta} {et~al.}(1993){Natta}, {Palla}, {Butner}, {Evans}, \&
  {Harvey}}]{1993ApJ...406..674N}
{Natta}, A., {Palla}, F., {Butner}, H.~M., {Evans}, II, N.~J., \& {Harvey},
  P.~M. 1993, \apj, 406, 674

\bibitem[{{Neuh{\"a}user} \& {Forbrich}(2008)}]{2008hsf2.book..735N}
{Neuh{\"a}user}, R. \& {Forbrich}, J. 2008, {The Corona Australis Star Forming
  Region}, ed. B.~{Reipurth}, 735

\bibitem[{{Pavlov} {et~al.}(2008){Pavlov}, {M{\"o}ller-Nilsson}, {Feldt},
  {Henning}, {Beuzit}, \& {Mouillet}}]{2008SPIE.7019E..39P}
{Pavlov}, A., {M{\"o}ller-Nilsson}, O., {Feldt}, M., {et~al.} 2008, in Society
  of Photo-Optical Instrumentation Engineers (SPIE) Conference Series, Vol.
  7019, Society of Photo-Optical Instrumentation Engineers (SPIE) Conference
  Series, 39

\bibitem[{{Pecaut} \& {Mamajek}(2013)}]{2013ApJS..208....9P}
{Pecaut}, M.~J. \& {Mamajek}, E.~E. 2013, \apjs, 208, 9

\bibitem[{{Percy} {et~al.}(2010){Percy}, {Grynko}, {Seneviratne}, \&
  {Herbst}}]{2010PASP..122..753P}
{Percy}, J.~R., {Grynko}, S., {Seneviratne}, R., \& {Herbst}, W. 2010, \pasp,
  122, 753

\bibitem[{{Racine} {et~al.}(1999){Racine}, {Walker}, {Nadeau}, {Doyon}, \&
  {Marois}}]{1999PASP..111..587R}
{Racine}, R., {Walker}, G.~A.~H., {Nadeau}, D., {Doyon}, R., \& {Marois}, C.
  1999, \pasp, 111, 587

\bibitem[{{Rayner} {et~al.}(2009){Rayner}, {Cushing}, \&
  {Vacca}}]{2009ApJS..185..289R}
{Rayner}, J.~T., {Cushing}, M.~C., \& {Vacca}, W.~D. 2009, \apjs, 185, 289

\bibitem[{{Rigliaco} {et~al.}(2012){Rigliaco}, {Natta}, {Testi}, {Randich},
  {Alcal{\`a}}, {Covino}, \& {Stelzer}}]{2012A&A...548A..56R}
{Rigliaco}, E., {Natta}, A., {Testi}, L., {et~al.} 2012, \aap, 548, A56

\bibitem[{{Sicilia-Aguilar} {et~al.}(2011){Sicilia-Aguilar}, {Henning},
  {Kainulainen}, \& {Roccatagliata}}]{2011ApJ...736..137S}
{Sicilia-Aguilar}, A., {Henning}, T., {Kainulainen}, J., \& {Roccatagliata}, V.
  2011, \apj, 736, 137

\bibitem[{{Sivaramakrishnan} \& {Oppenheimer}(2006)}]{2006ApJ...647..620S}
{Sivaramakrishnan}, A. \& {Oppenheimer}, B.~R. 2006, \apj, 647, 620

\bibitem[{{Soummer} {et~al.}(2012){Soummer}, {Pueyo}, \&
  {Larkin}}]{2012ApJ...755L..28S}
{Soummer}, R., {Pueyo}, L., \& {Larkin}, J. 2012, \apjl, 755, L28

\bibitem[{{Takami} {et~al.}(2003){Takami}, {Bailey}, \&
  {Chrysostomou}}]{2003A&A...397..675T}
{Takami}, M., {Bailey}, J., \& {Chrysostomou}, A. 2003, \aap, 397, 675

\bibitem[{{Taylor} \& {Storey}(1984)}]{1984MNRAS.209P...5T}
{Taylor}, K.~N.~R. \& {Storey}, J.~W.~V. 1984, \mnras, 209, 5P

\bibitem[{{The}(1994)}]{1994ASPC...62...23T}
{The}, P.~S. 1994, in Astronomical Society of the Pacific Conference Series,
  Vol.~62, The Nature and Evolutionary Status of Herbig Ae/Be Stars, ed. P.~S.
  {The}, M.~R. {Perez}, \& E.~P.~J. {van den Heuvel}, 23

\bibitem[{{Vigan} {et~al.}(2010){Vigan}, {Moutou}, {Langlois}, {Allard},
  {Boccaletti}, {Carbillet}, {Mouillet}, \& {Smith}}]{2010MNRAS.407...71V}
{Vigan}, A., {Moutou}, C., {Langlois}, M., {et~al.} 2010, \mnras, 407, 71

\bibitem[{{Ward-Thompson} {et~al.}(1985){Ward-Thompson}, {Warren-Smith},
  {Scarrott}, \& {Wolstencroft}}]{1985MNRAS.215..537W}
{Ward-Thompson}, D., {Warren-Smith}, R.~F., {Scarrott}, S.~M., \&
  {Wolstencroft}, R.~D. 1985, \mnras, 215, 537

\bibitem[{{Wilking} {et~al.}(1997){Wilking}, {McCaughrean}, {Burton}, {Giblin},
  {Rayner}, \& {Zinnecker}}]{1997AJ....114.2029W}
{Wilking}, B.~A., {McCaughrean}, M.~J., {Burton}, M.~G., {et~al.} 1997, \aj,
  114, 2029

\bibitem[{{Zechmeister} \& {K{\"u}rster}(2009)}]{2009A&A...496..577Z}
{Zechmeister}, M. \& {K{\"u}rster}, M. 2009, \aap, 496, 577

\bibitem[{{Zurlo} {et~al.}(2018){Zurlo}, {Mesa}, {Desidera}, {Messina},
  {Gratton}, {Moutou}, {Beuzit}, {Biller}, {Boccaletti}, {Bonavita},
  {Bonnefoy}, {Bhowmik}, {Brandner}, {Buenzli}, {Chauvin}, {Cudel}, {D'Orazi},
  {Feldt}, {Hagelberg}, {Janson}, {Lagrange}, {Langlois}, {Lannier}, {Lavie},
  {Lazzoni}, {Maire}, {Meyer}, {Mouillet}, {Peretti}, {Perrot}, {Potiron},
  {Salter}, {Schmidt}, {Sissa}, {Vigan}, {Delboulb{\'e}}, {Petit}, {Ramos},
  {Rigal}, \& {Rochat}}]{2018MNRAS.480...35Z}
{Zurlo}, A., {Mesa}, D., {Desidera}, S., {et~al.} 2018, \mnras, 480, 35

\bibitem[{{Zurlo} {et~al.}(2013){Zurlo}, {Vigan}, {Hagelberg}, {Desidera},
  {Chauvin}, {Almenara}, {Biazzo}, {Bonnefoy}, {Carson}, {Covino}, {Delorme},
  {D'Orazi}, {Gratton}, {Mesa}, {Messina}, {Moutou}, {Segransan}, {Turatto},
  {Udry}, \& {Wildi}}]{2013A&A...554A..21Z}
{Zurlo}, A., {Vigan}, A., {Hagelberg}, J., {et~al.} 2013, \aap, 554, A21

\bibitem[{{Zurlo} {et~al.}(2014){Zurlo}, {Vigan}, {Mesa}, {Gratton}, {Moutou},
  {Langlois}, {Claudi}, {Pueyo}, {Boccaletti}, {Baruffolo}, {Beuzit},
  {Costille}, {Desidera}, {Dohlen}, {Feldt}, {Fusco}, {Henning}, {Kasper},
  {Martinez}, {Moeller-Nilsson}, {Mouillet}, {Pavlov}, {Puget}, {Sauvage},
  {Turatto}, {Udry}, {Vakili}, {Waters}, \& {Wildi}}]{zurlo2014}
{Zurlo}, A., {Vigan}, A., {Mesa}, D., {et~al.} 2014, \aap, 572, A85

\end{thebibliography}

\end{document}